# ALMA-IMF IX: Catalog and Physical Properties of 315 SiO Outflow Candidates in 15 Massive Protoclusters

A. P. M. Towner,[1] A. Ginsburg,[2] P. Dell'Ova,[3] A. Gusdorf,[3,4] S. Bontemps,[5] T. Csengeri,[5] R. Galván-Madrid,[6]
F. K. Louvet,[7] F. Motte,[7] P. Sanhueza,[8,9] A. M. Stutz,[10] J. Bally,[11] T. Baug,[12] H.-R. V. Chen,[13]
N. Cunningham,[7] M. Fernández-López,[14] H.-L. Liu,[15] X. Lu,[16] T. Nony,[17] M. Valeille-Manet,[5] B. Wu,[8,18]
R. H. Álvarez-Gutiérrez,[10] M. Bonfand,[19] J. Di Francesco,[20] Q. Nguyen-Luong,[21] F. Olguin,[13] AND
A. P. Whitworth[22]

[1] University of Arizona Department of Astronomy and Steward Observatory, 933 North Cherry Ave., Tucson, AZ 85721, USA
[2] Department of Astronomy, University of Florida, PO Box 112055, USA
[3] Laboratoire de Physique de l'École Normale Supérieure, ENS, Université PSL, CNRS, Sorbonne Université, Université de Paris, Paris, France
[4] Observatoire de Paris, PSL University, Sorbonne Université, LERMA, 75014, Paris, France
[5] Laboratoire d'astrophysique de Bordeaux, Univ. Bordeaux, CNRS, B18N, allée Geoffroy Saint-Hilaire, 33615 Pessac, France
[6] Instituto de Radioastronomía y Astrofísica, Universidad Nacional Autónoma de México, Morelia, Michoacán 58089, México
[7] Univ. Grenoble Alpes, CNRS, IPAG, 38000 Grenoble, France
[8] National Astronomical Observatory of Japan, National Institutes of Natural Sciences, 2-21-1 Osawa, Mitaka, Tokyo 181-8588, Japan
[9] Department of Astronomical Science, SOKENDAI (The Graduate University for Advanced Studies), 2-21-1 Osawa, Mitaka, Tokyo 181-8588, Japan
[10] Departamento de Astronomía, Universidad de Concepción, Casilla 160-C, Concepción, Chile
[11] Department of Astrophysical and Planetary Sciences, University of Colorado, Boulder, Colorado 80389, USA
[12] S. N. Bose National Centre for Basic Sciences, Block JD, Sector III, Salt Lake, Kolkata 700106, India
[13] Institute of Astronomy, National Tsing Hua University, Hsinchu 30013, Taiwan
[14] Instituto Argentino de Radioastronomía (CCT-La Plata, CONICET; CICPBA), C.C. No. 5, 1894, Villa Elisa, Buenos Aires, Argentina
[15] Department of Astronomy, Yunnan University, Kunming, 650091, PR China
[16] Shanghai Astronomical Observatory, Chinese Academy of Sciences, 80 Nandan Road, Shanghai 200030, People's Republic of China
[17] Instituto de Radioastronomía y Astrofísica, Universidad Nacional Autónoma de México, Morelia, Michoacán 58089, México
[18] NVIDIA Research, 2788 San Tomas Expy, Santa Clara, CA 95051, USA
[19] Department of Astronomy, University of Virginia, 530 McCormick Rd., Charlottesville, VA 22903
[20] Herzberg Astronomy and Astrophysics Research Centre, National Research Council of Canada, 5071 West Saanich Road, Victoria, BC V9E 2E7 Canada
[21] Department of Computer Science, Mathematics & Environmental Science, The American University of Paris, PL111, 2 bis, passage Landrieu 75007 Paris, France
[22] School of Physics and Astronomy, Cardiff University, N/3.27, Queen's Buildings - North Building, 5 The Parade, Newport Road, Cardiff, CF24 3AA

## ABSTRACT

We present a catalog of 315 protostellar outflow candidates detected in SiO $J = 5 - 4$ in the ALMA-IMF Large Program, observed with $\sim$2000 au spatial resolution, 0.339 km s$^{-1}$ velocity resolution, and 2-12 mJy beam$^{-1}$ (0.18-0.8 K) sensitivity. We find median outflow masses, momenta, and kinetic energies of $\sim 0.3$ M$_\odot$, 4 M$_\odot$ km s$^{-1}$, and $10^{45}$ erg, respectively. Median outflow lifetimes are 6,000 years, yielding median mass, momentum, and energy rates of $\dot{M} = 10^{-4.4}$ M$_\odot$ yr$^{-1}$, $\dot{P} = 10^{-3.2}$ M$_\odot$ km s$^{-1}$ yr$^{-1}$, and $\dot{E} = 1$ L$_\odot$. We analyze these outflow properties in the aggregate in each field. We find correlations between field-aggregated SiO outflow properties and total mass in cores ($\sim$3–5$\sigma$), and no correlations above 3$\sigma$ with clump mass, clump luminosity, or clump luminosity-to-mass ratio. We perform a linear regression analysis and find that the correlation between field-aggregated outflow mass and total clump mass — which has been previously described in the literature — may actually be mediated by the relationship between outflow mass and total mass in cores. We also find that the most massive SiO outflow in each field is typically responsible for only 15-30% of the total outflow mass (60% upper limit). Our data agree well with the established mechanical force−bolometric luminosity relationship in the literature, and our data extend this relationship up to L $\geq 10^6$ L$_\odot$ and $\dot{P} \geq 1$ M$_\odot$ km s$^{-1}$ yr$^{-1}$. Our lack of correlation with clump $L/M$ is inconsistent with models of protocluster formation in which all protostars start forming at the same time.

## 1. INTRODUCTION

Outflows are observed from accreting stars of all masses, but the relative impact of outflows from low- and high-mass stars in clustered environments is still debated (Krumholz et al. 2014, and references therein). Part of this debate stems from the historic limitation that high-mass star-forming regions were typically observed with coarser spatial resolution than low-mass regions, due to their larger distances from Earth. As a result, we have not been able to probe the full population of individual protostars, and the protostellar feedback they produce, in a statistical sample of massive protoclusters. Protostellar outflows are observed to occur at all stages of protocluster evolution (Bally 2016; Svoboda et al. 2019; Nony et al. 2020, and references therein), and at early times are assumed to be the strongest type of protostellar feedback (Krumholz et al. 2014, and references therein). This makes them an excellent tool for probing protostellar populations, and the relative impact of protostars of different masses on the protocluster overall, across a range of protocluster evolutionary states.

Perhaps the best-known molecular tracer of the high-velocity component of protostellar outflows is silicon monoxide (SiO). This molecule is often found to be coincident with high-velocity shocks in star-forming regions (Bally 2016; Dutta et al. 2020; Morii et al. 2021). SiO is expected to trace shocks particularly well due to the high collision velocities ($\gtrsim$ 25 km s$^{-1}$) required to release Si-bearing material from dust grain cores (Gusdorf et al. 2008; Schilke et al. 1997). It has been used by numerous teams to study outflows from protostars spanning a wide range of masses, from low-mass samples (Dutta et al. 2020; Lee 2020, and references therein) to high-mass young stellar objects (López-Sepulcre et al. 2011; Sánchez-Monge et al. 2013; Csengeri et al. 2016; Liu et al. 2021; Lu et al. 2021; Liu et al. 2022). SiO J=5−4 typically has a higher detection rate in high-mass star-forming regions (e.g. Nony et al. 2020; Li et al. 2020) than in low-mass (Dutta et al. 2020), likely as a reflection of the high critical densities required to excite this transition ($10^5 - 10^6$ cm$^{-3}$, Gusdorf et al. 2008; Leurini et al. 2014). SiO outflows have been detected even in massive protoclusters at very early stages (Svoboda et al. 2019; Li et al. 2020), though with lower detection rates than are found in more-evolved regions (Csengeri et al. 2016; Nony et al. 2020).

A number of studies have examined outflow physical properties, and in particular outflow mechanical force

(momentum per time), across a range of protostellar masses, e.g. Bontemps et al. (1996) for 45 protostars in the low-mass regime, (Duarte-Cabral et al. 2013) for a sample of 9 individual high-mass protostars, and Maud et al. (2015) for 99 high-mass protoclusters at clump-scale ($\gtrsim$0.1 pc) resolution. However, a major limitation of outflow population studies in the high-mass regime is the embedded nature of and large distances to most high-mass protoclusters. This makes separating individual continuum cores and outflows difficult in these clustered environments. To date, many of the largest surveys of protostellar outflows in massive star-forming regions were limited by small-number statistics (Duarte-Cabral et al. 2013), have >10″ angular resolution (Liu et al. 2022; Maud et al. 2015), or had low detection rates of SiO outflows specifically, possibly due to the early evolutionary stages of the targets (Li et al. 2020; Svoboda et al. 2019; de Villiers et al. 2014; López-Sepulcre et al. 2009). Consequently, many studies have probed SiO-detected protostellar outflow properties at the scale of the whole protocluster and its host clump ($\sim$0.1−1 pc) rather than at the scale of individual star-forming cores ($\lesssim$0.01 pc).

This limitation has led to observationally-derived correlations between various clump properties ($M_{clump}$, $L_{bol,\,clump}$) and outflow properties (total mass in outflows, outflow mechanical force, etc; see e.g. Beuther et al. 2002b; Csengeri et al. 2016; Liu et al. 2022), but not protostellar-scale correlations. It has also led to the assumption that the most massive core produces the most massive outflow, which in turn dominates the total mass in outflows in the protocluster (Maud et al. 2015). Because star-forming cores in massive star-forming regions are often clustered and outflows from adjacent sources can overlap along the line of sight, confirming the origins of these correlations and the accuracy of these assumptions requires protostellar-scale line observations in order to characterize each outflow individually.

We present the first comprehensive catalog of 315 SiO-identified protostellar outflows in the 15 massive protoclusters targeted by the ALMA-IMF Large Program. ALMA-IMF (Motte et al. 2022) seeks to explore the shape and evolution of the Core Mass Function (CMF) by observing a sample of 15 massive protoclusters using the Atacama Large Millimeter/submillimeter Array (ALMA). The protoclusters were observed in both line and continuum emission at 1.3 mm (Band 6) and 3 mm (Band 3) with $\sim$2000 au resolution. The ALMA-IMF fields were selected to span a range of evolutionary states (Young, Intermediate, and Evolved) in order to explore variation of the CMF with time in a statistical sample of continuum cores. In this paper, we analyze the SiO emission in these 15 protoclusters. In order to perform an unbiased search for SiO emission, we search the SiO images directly rather than starting from the locations of known continuum sources (Nony et al. 2020, 2023). This large, homogeneous sample of outflow candidates will serve as a comprehensive resource for follow-up studies of the individual outflows and overall outflow populations in these fields.

In Table 1, we show basic information for the 15 protoclusters in our sample. In § 2 we describe the observational details and image properties of the SiO data used in this work, which have now been released publicly. In § 3 we present the catalog, describe the procedure used to identify and confirm or reject outflow candidates, and derive the physical properties of each outflow candidate (mass, momentum, energy, outflow lifetime, mass rate, mechanical force, and energy rate). In § 4 we compare our candidates to similar samples in the literature, and discuss our derived correlations between field-aggregated outflow properties and clump properties ($M_{clump}$, $M_{cores}$, $L_{bol}$, and $L/M$). We also discuss the dominance (or lack thereof) of the strongest outflow in each field over field-aggregated outflow properties, and discuss the possible origins of the known correlation between clump mass and total mass in outflows (e.g. Beuther et al. 2002b). In § 5 we present our summary and conclusions.

## 2. OBSERVATIONS

The SiO J=5−4 data presented herein were taken as part of the ALMA-IMF Large Program (2017.1.01355.L, PIs: Motte, Ginsburg, Louvet, Sanhueza), with the exception of the SiO observations for W43-MM1, which were taken as part of the pilot program 2013.1.01365.S (Nony et al. 2020). The ALMA-IMF Large Program targets were observed in Band 6 ($\sim$216-234 GHz, $\sim$1.3 mm) and Band 3 ($\sim$91-106 GHz, $\sim$3 mm), with matching linear spatial resolution ($\lesssim$2000 au) for all fields and in both bands. The more distant targets (see Table 1) were observed with two 12m configurations and the ACA (7m+TP), while closer targets were observed with only one 12m configuration and the ACA. The SiO J=5−4 line has a rest frequency of 217.10498 GHz. This line is located in spectral window 1 (spw1) in the ALMA-IMF Large Program tuning, with channel widths of $\Delta$v = 0.339 km s$^{-1}$. Details of the tuning setup, including the array configurations, bandwidth, spectral resolution, and main spectral lines for each spectral window can be found in Motte et al. (2022) and Ginsburg et al. (2022). Additional details of the data reduction and `tclean` imaging parameters for the line data specifically can be found in Cunningham et al. (2023). In this work, we examine only those data taken with the 12m array. These line cubes can be found at https://www.almaimf.com/data.html. The combined, 12m+7m+TP data from the ALMA-IMF Large Program will be released in future planned publications (Stutz et al. in prep, Álvarez-Gutiérrez et al in prep, Sandoval et al. in prep).

All line cubes presented herein were corrected for the "Jorsater & van Moorsel effect" (or "JvM effect"), which arises because the size of the CLEAN beam, which is convolved with the CLEAN model points, is different from that of the dirty beam contained in the residual image. In order to create an image with self-consistent units in both the modeled and residual emission, these two different beam sizes must be accounted for (Jorsater & van Moorsel 1995). We apply the "JvM correction" to our data using the method described in Czekala et al. (2021), in which the residual image is scaled by the ratio of the CLEAN and dirty beam volumes before the restored image is created. The cubes were then continuum-subtracted in the image plane using the `statcont` task in python (Sánchez-Monge et al. 2018). We then apply a primary beam correction to each cube. For the remainder of the paper, we use the JvM-corrected, primary beam-corrected, continuum-subtracted line cubes for our analysis. Table 2 shows relevant image and statistical information for each resulting spw1 cube.

## 3. RESULTS

### 3.1. *Preparation of the SiO Cubes*

Starting from the fully-processed cubes for spw1, we use the SpectralCube package[1] to create subcubes for each field extending from $V_{LSR} - 95$ km s$^{-1}$ to $V_{LSR} + 95$ km s$^{-1}$, using a rest frequency of $\nu_{SiO} = 217.10498$ GHz. This is the maximum velocity coverage common to all 15 fields in the sample based on the tuning for each target. We also created $V_{LSR} \pm 95$ km s$^{-1}$ cutouts from the spw1 primary beam files produced by our imaging pipeline (see Ginsburg et al. 2022; Cunningham et al. 2023).

---

[1] https://github.com/radio-astro-tools/spectral-cube

**Table 1.** ALMA-IMF Protocluster Properties

| Field | Distance[a] | $V_{LSR}$[a] | $\Delta V_{LSR}$[b] | $M_{cores}$[c] | $M_{clump}$[d] | $L_{clump}$[d] | L/M[e] | Evo.[f] |
|---|---|---|---|---|---|---|---|---|
| | (kpc) | (km s$^{-1}$) | (km s$^{-1}$) | (M$_\odot$) | (10$^3$M$_\odot$) | (10$^3$L$_\odot$) | (L$_\odot$/M$_\odot$) | State |
| G008.67 | $3.4 \pm 0.3$ | +37.6 | 7.3 | 104 (3) | 5 (1) | 82 (10) | 16 | I |
| G010.62 | $4.9 \pm 0.5$ | $-2$ | 10.1 | 189 (5) | 12 (2) | 430 (100) | 36 | E |
| G012.80 | $2.4 \pm 0.2$ | +37 | 7.4 | 207 (4) | 7 (1) | 310 (50) | 44 | E |
| G327.29 | $2.5 \pm 0.5$ | $-45$ | 5.7 | 428 (3) | 10 (4) | 100 (40) | 10 | Y |
| G328.25 | $2.5 \pm 0.5$ | $-43$ | 2.5 | 38.7 (0.7) | 2 (1) | 46 (20) | 23 | Y |
| G333.60 | $4.2 \pm 0.7$ | $-47$ | 10.4 | 444 (4) | 19 (10) | 1500 (500) | 79 | E |
| G337.92 | $2.7 \pm 0.7$ | $-40$ | 3.9 | 133 (4) | 5 (2) | 120 (50) | 24 | Y |
| G338.93 | $3.9 \pm 1.0$ | $-62$ | 7.7 | 250 (2) | 6 (3) | 100 (50) | 17 | Y |
| G351.77 | $2.0 \pm 0.7$ | $-3$ | 6.0 | 144 (3) | 2 (1) | 100 (60) | 50 | I |
| G353.41 | $2.0 \pm 0.7$ | $-17$ | 7.6 | 142 (3) | 3 (2) | 87 (50) | 29 | I |
| W43—MM1 | $5.5 \pm 0.4$ | +97 | 7.0 | 634 (6) | 17 (2) | 210 (30) | 12 | Y |
| W43—MM2 | $5.5 \pm 0.4$ | +97 | 4.7 | 298 (2) | 25 (4) | 170 (20) | 7 | Y |
| W43—MM3 | $5.5 \pm 0.4$ | +97 | 4.6 | 104 (2) | 13 (2) | 140 (20) | 11 | I |
| W51—E | $5.4 \pm 0.3$ | +55 | 11.7 | 830 (14) | 61 (10) | 1000 (100) | 16 | I |
| W51—IRS2 | $5.4 \pm 0.3$ | +55 | 13.7 | 905 (7) | 29 (3) | 1800 (200) | 62 | E |

[a] From Motte et al. (2022), Table 1.

[b] The total variation in $V_{LSR}$ within the clump, as derived from single-component fits to DCN line emission associated with continuum cores. See Cunningham et al. (2023), Table 4, Column 5 and associated text for the DCN fitting procedure and results. See Louvet et al. (2023, submitted) for the catalog of continuum cores.

[c] From Louvet et al. (2023, submitted). This is the 1.3 mm continuum-derived mass of all cores within the field mosaic identified with the **getsf** algorithm using the **cleanest** (line-free) images, taking contamination from free-free emission into account. See Ginsburg et al. (2022) for details of the **cleanest** images, and Men'shchikov (2021) for details of the **getsf** tool.

[d] From Dell'Ova et al. (2023, in prep), derived using the PPMAP tool using 3.6 $\mu$m through 1.3 mm images. See Marsh et al. (2015) for details of the PPMAP tool.

[e] The ratio of $L_{clump}$ to $M_{clump}$ in the preceding two columns.

[f] The overall evolutionary state of each protocluster, taken from Motte et al. (2022), Table 4, Column 8. Y = Young, I = Intermediate, E = Evolved.

Our noise levels typically vary by 4−8% between channels for a given field (columns 8 and 9 of Table 2), but can vary by as much as 11−20% (G338.93, G351.77) or as little as 2.8% (W43-MM1). We take our noise in a single channel to be $\sigma = 1.4826 \times$MAD[2]. The noise is measured within an emission-free polygonal region near the center of each field of view; the same emission-free region is used for all channels in a given cube. The largest noise variations in our data appear to be caused by imaging artifacts.

In order to use the most accurate noise level for each outflow candidate, we used SpectralCube to create "noise cubes" whose values vary by both channel and pixel. We take the noise in each channel ($\sigma_{channel}$) and, at each pixel location, divide $\sigma_{channel}$ by the value of the primary beam correction at that pixel. This has little effect on $\sigma$ near the center of the image (where the primary beam response is ∼1) but will increase $\sigma$ towards the edges of each mosaic. In this way, we created noise cubes in which the noise level varies with frequency according to $\sigma$ measured for each channel, and varies spatially according to the effects of the primary beam.

---

[2] MAD is the median absolute deviation from the median within a line-free region in each channel. The factor of 1.4826 relates MAD and standard deviation for a Gaussian distribution; the term "1.4826×MAD" is sometimes called the scaled MAD.

**Table 2.** ALMA-IMF SiO Line Cube Properties

| Field | RA[a] | Dec[a] | Synthesized Beam | | Pixel | | Median $\sigma$[b] | Min, Max $\sigma$[b] |
|---|---|---|---|---|---|---|---|---|
| | ($^h$ $^m$ $^s$) | ($^\circ$ $'$ $''$) | ($'' \times ''$) | ($^\circ$) | Size | (K) | (mJy beam$^{-1}$) | (mJy beam$^{-1}$) |
| G008.67 | 18:06:21.12 | −21:37:16.7 | $0.88 \times 0.72$ | −81 | $0\rlap{.}''12$ | 0.37 | 8.86 | 8.55, 9.29 |
| G010.62 | 18:10:28.80 | −19:55:48.3 | $0.68 \times 0.53$ | −73 | $0\rlap{.}''11$ | 0.22 | 3.00 | 2.94, 3.10 |
| G012.80 | 18:14:13.37 | −17:55:45.2 | $1.29 \times 0.88$ | 77 | $0\rlap{.}''18$ | 0.31 | 13.89 | 13.08, 14.99 |
| G327.29 | 15:53:08.13 | −54:37:08.6 | $0.82 \times 0.75$ | −56 | $0\rlap{.}''12$ | 0.51 | 12.09 | 11.64, 12.60 |
| G328.25 | 15:57:59.68 | −53:57:59.8 | $0.74 \times 0.58$ | −14 | $0\rlap{.}''12$ | 0.90 | 14.97 | 14.26, 15.57 |
| G333.60 | 16:22:09.36 | −50:05:59.2 | $0.75 \times 0.68$ | −36 | $0\rlap{.}''11$ | 0.20 | 3.94 | 3.81, 4.25 |
| G337.92 | 16:41:10.62 | −47:08:02.9 | $0.80 \times 0.66$ | −51 | $0\rlap{.}''11$ | 0.23 | 4.70 | 4.43, 4.98 |
| G338.93 | 16:40:34.42 | −45:41:40.6 | $0.77 \times 0.68$ | 82 | $0\rlap{.}''11$ | 0.23 | 4.65 | 4.25, 5.61 |
| G351.77 | 17:26:42.62 | −36:09:20.5 | $1.08 \times 0.83$ | 88 | $0\rlap{.}''17$ | 0.39 | 13.57 | 12.79, 15.12 |
| G353.41 | 17:30:26.28 | −34:41:49.7 | $1.13 \times 0.83$ | 86 | $0\rlap{.}''17$ | 0.43 | 15.58 | 14.19, 16.85 |
| W43−MM1 | 18:47:46.50 | −01:54:29.5 | $0.65 \times 0.47$ | −80 | $0\rlap{.}''08$ | 0.39 | 4.59 | 4.46, 4.72 |
| W43−MM2 | 18:47:36.61 | −02:00:51.7 | $0.62 \times 0.51$ | −85 | $0\rlap{.}''08$ | 0.20 | 2.41 | 2.24, 2.58 |
| W43−MM3 | 18:47:41.46 | −02:00:28.2 | $0.66 \times 0.57$ | 86 | $0\rlap{.}''11$ | 0.18 | 2.65 | 2.46, 2.86 |
| W51−E | 19:23:44.18 | +14:30:28.9 | $0.46 \times 0.35$ | 30 | $0\rlap{.}''08$ | 0.45 | 2.74 | 2.66, 2.85 |
| W51−IRS2 | 19:23:39.81 | +14:31:02.9 | $0.65 \times 0.59$ | −23 | $0\rlap{.}''11$ | 0.37 | 5.50 | 5.29, 5.74 |

[a] The ICRS coordinates of the reference position (center) of each mosaic, taken from the SiO line cube image headers.

[b] Median $\sigma$ for each cube, in both K and mJy beam$^{-1}$. In all cases, $\sigma = 1.4826 \times$ MAD, where MAD is the median absolute deviation from the median. $\sigma$ is measured for every channel in the image cube within an emission-free region near the center of the field. The emission-free region is the same for all channels in a given field. The median, minimum, and maximum $\sigma$ reported in these columns are calculated across all channels in the cube.

Using these 3-dimensional noise cubes, we masked each image cube spectrally and spatially at the $3\sigma$ level. The resulting maps still contain some spurious emission, as expected for a $3\sigma$ cutoff given our typical $\sim 10^9$ pixels in a cube. In order to remove this emission, we used the `scipy.ndimage` package to perform 3-dimensional binary erosion (1 iteration) and binary dilation (2 iterations) on each mask[3]. This procedure is equivalent to requiring that emission be present above the $3\sigma$ level in both spectral ($\geq 3$ consecutive channels) and spatial ($\geq 3$ pixels across) dimensions. The erosion/dilation step successfully removed nearly all of the spurious emission from our data cubes with minimal loss of true signal. Using these masked cubes, we created integrated-intensity (moment 0), intensity-weighted ve-locity (moment 1), intensity-weighted variance (moment 2), linewidth ($\sqrt{moment\,2}$), and maximum-intensity maps for each field.

### 3.2. Outflow Candidate Identification Procedure

We performed an initial search for protostellar outflows using these maps and the unmasked line cubes. All examinations, including the initial inspection discussed above, were performed in the Cube Analysis and Rendering Tool for Astronomy (CARTA; Comrie et al. 2021). We identified candidates first by eye based on linear morphology in any map plus $V - V_{LSR} > 5$ km s$^{-1}$ in the moment 1 maps or linewidth $> 10$ km s$^{-1}$ in the linewidth maps. We then examined the line cubes directly to ensure that no regions with emission of $V - V_{LSR} > 5$ km s$^{-1}$ were missed in the moment-map examination. We do not require a continuum driving source to be positively identified in order to list a candidate, and we do not report specific driving sources for any outflows in this paper. However, the presence of a continuum source coincident with an outflow may increase our confidence that a particular candidate is indeed an outflow, or that a red- and blue-shifted lobe share a driving source (see § 3.3, below).

---

[3] For G333.60 only, we use 2 iterations for erosion and 3 for dilation. This procedure is equivalent to requiring $>3\sigma$ emission in $\geq 5$ consecutive channels and $\geq 5$ pixels across for this field only. This increase in erosion/dilation iterations is due to persistent cleaning artifacts in the G333.60 line cube.

After initial identification, we drew a custom polygonal region around each candidate using its $3\sigma$ contours in the masked moment maps. We then modified (expanded) this region as appropriate based on by-eye examination of each channel in the line cube, to make sure no relevant emission was missed (e.g. faint or high-velocity). Each candidate's emission was then integrated spatially over this custom polygonal aperture in each channel to create an aperture-integrated spectrum, which was then examined by eye. We also generated a position-velocity (pv) diagram along each candidate's longest axis using the radio-astro-tools package PV Extractor[4]. Finally, each candidate's overall structure was examined directly in the image cube channel-by-channel. Candidates were confirmed or rejected, and polygonal apertures modified, based on this final, spatial and kinematic examination.

For each outflow, we produce a summary figure showing its integrated intensity, velocity-weighted intensity, linewidth, and maximum-value maps, along with the aperture-integrated spectrum, PV diagram, and 1.3 mm continuum emission map. Examples are shown in Figures 1 and 2. The candidate shown in Figure 1 is a typical, symmetric bipolar outflow, while that shown in Figure 2 is bipolar but asymmetric, with a significantly larger, brighter, and higher-velocity red-shifted lobe than blue-shifted lobe. Detailed examination of individual candidates or fields is beyond the scope of this work.

Summary figures for all candidates can be viewed on Zenodo: doi:10.5281/zenodo.8350595. An initial list of outflow candidates was made available internally within the collaboration, with voluntary feedback requested. Following review and evaluation by 14 members, this initial list of candidates was refined.

### 3.3. The Catalog

In total, we detect 315 outflow candidates across 15 fields. Of the 315 candidates, 39 are classified as bipolar, 147 as blue monopolar, and 129 as red monopolar[5]. We find a median of 17 candidates per field; the minimum number of candidates is 3 (G328.25), and the maximum is 47 (W51-IRS2). Our full catalog is presented in a machine-readable format in the online Journal; a representative example of the catalog is shown in Table 3. The full catalog can also be found in ESCV format on

Zenodo: doi:10.5281/zenodo.8350595. For each outflow candidate, we report its approximate center position, its color, the total velocity range over which high-velocity emission is detected, the velocity at which the aperture-integrated spectrum peaks, the peak intensity of the aperture-integrated spectrum, the aperture- and velocity-integrated flux density, and our classification for that candidate.

The center positions are calculated as the center of the bounding box encompassing the polygonal aperture used for each outflow candidate. That is, the center RA for each candidate is the average of the minimum and maximum RA values in that candidate's polygonal aperture, and the center Dec is the average of the minimum and maximum Dec values. Candidates with colors listed as 'red+blue' are bipolar. Candidates listed with just a single color ('red' or 'blue') are either monopolar or, if potentially bipolar, a counterpart cannot be definitively identified from among multiple possibilities. The latter is especially common in regions with significant outflow activity and/or a high local number density of cores. Candidates are only listed as 'red+blue' (bipolar) if the same 1.3 mm continuum driving source can be associated with both lobes with high confidence.

We identify velocity ranges by eye for each candidate based on its aperture-integrated spectrum, its position-velocity diagram, and the unmasked line cube. In general, velocity ranges for each outflow exclude $V_{LSR,\,candidate} \pm 5$ km s$^{-1}$ based on line shape, where $V_{LSR,\,candidate}$ is the standard of rest velocity assessed locally for each candidate. We assess $V_{LSR,\,candidate}$ locally because the clump $V_{LSR}$ can vary across the field (see Table 1). In some rare cases, we exclude more or less of the velocity range around $V_{LSR,candidate}$, based on line shape in the integrated spectrum and channel-by-channel by-eye examination of candidate morphology. We derive physical properties for each candidate in § 3.4. Although we restrict our search for outflows to velocities of $V_{LSR} \pm 95$ km s$^{-1}$, the aperture-integrated spectra of five candidates (W43-MM2 Candidate #24; W51-E Candidates #9, #19, and #20; W51-IRS2 Candidate #16) show emission at even higher velocities. We use the full spw1 cubes for our analyses of these candidates only. For W43-MM1 Candidate #24, W51-E Candidate #20, and W51-IRS Candidate #16, the full outflow spectrum is cut off even in the spw1 cube, so the reported velocity ranges for those candidates should be considered lower limits.

We use the following three classifications: 1) likely, 2) possible, and 3) complex or cluster. "Likely" candidates are those we consider significantly likely to be protostellar outflows, based on their brightness, morphology, aperture-integrated spectrum, and structure in the PV diagram. Most of the candidates which have bright continuum sources in or near the polygonal aperture fall into this category. "Possible" candidates are those we consider likely or probable outflow candidates, but either their brightness, morphology, spectral structure, or PV structure is not quite definitive enough to place them in category #1. "Complex or cluster" candidates are those which clearly exhibit high-velocity emission but either a) do not display the typical morphology of a protostellar outflow or b) appear to be blended emission from multiple outflows and individual driving sources cannot be identified. In total, 129 candidates are classified as "likely," 180 are classified as "possible," and 6 are classified as "complex or cluster."

---

[5] Counting the red and blue lobes in bipolar candidates separately, we have a total of 354 outflow candidates (186 blue, 168 red).

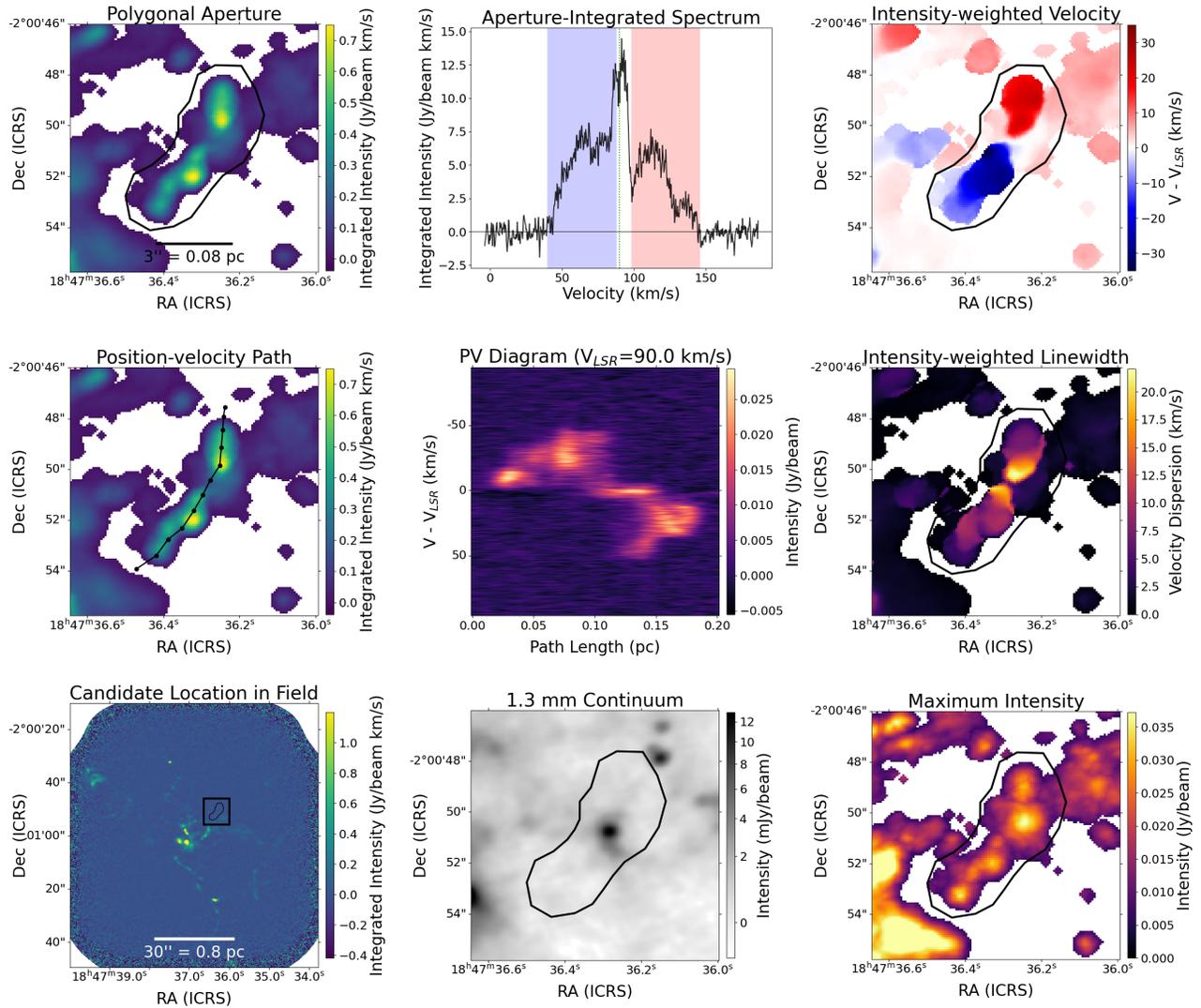

**Figure 1.** Example summary figure of a symmetric bipolar outflow from our sample (Candidate #8 in field W43-MM2). The top left panel shows the integrated-intensity (moment 0) map with the candidate's polygonal aperture overlaid in black. The top middle panel shows the candidate's spectrum integrated over the polygonal aperture, with red- and blue-lobe velocity intervals shaded red and blue, respectively. Velocities are absolute, and the field $V_{LSR}$ is shown as a vertical green dotted line. The center left panel again shows the candidate's integrated-intensity map, this time with the position-velocity diagram path overlaid in black. The center panel shows the position-velocity diagram for the candidate, where the y-axis shows velocity relative to field $V_{LSR}$. The bottom left panel shows the integrated-intensity map for the full field of view for W43-MM2, with the candidate's polygonal aperture overlaid in black and boxed. The center bottom panel shows the 1.3 mm continuum image at the same scale as the top left panel, with the polygonal aperture overlaid in black. The top, middle, and bottom panels in the right-hand column show the intensity-weighted velocity (moment 1), intensity-weighted linewidth ($\sqrt{\text{moment 2}}$), and maximum intensity maps, respectively. All three panels are at the same scale as the top left panel, and the candidate's polygonal aperture is overlaid in black in each. The colorbar for the top right panel shows velocity relative to $V_{LSR}$. The complete figure set (315 images) is available in the online Journal.

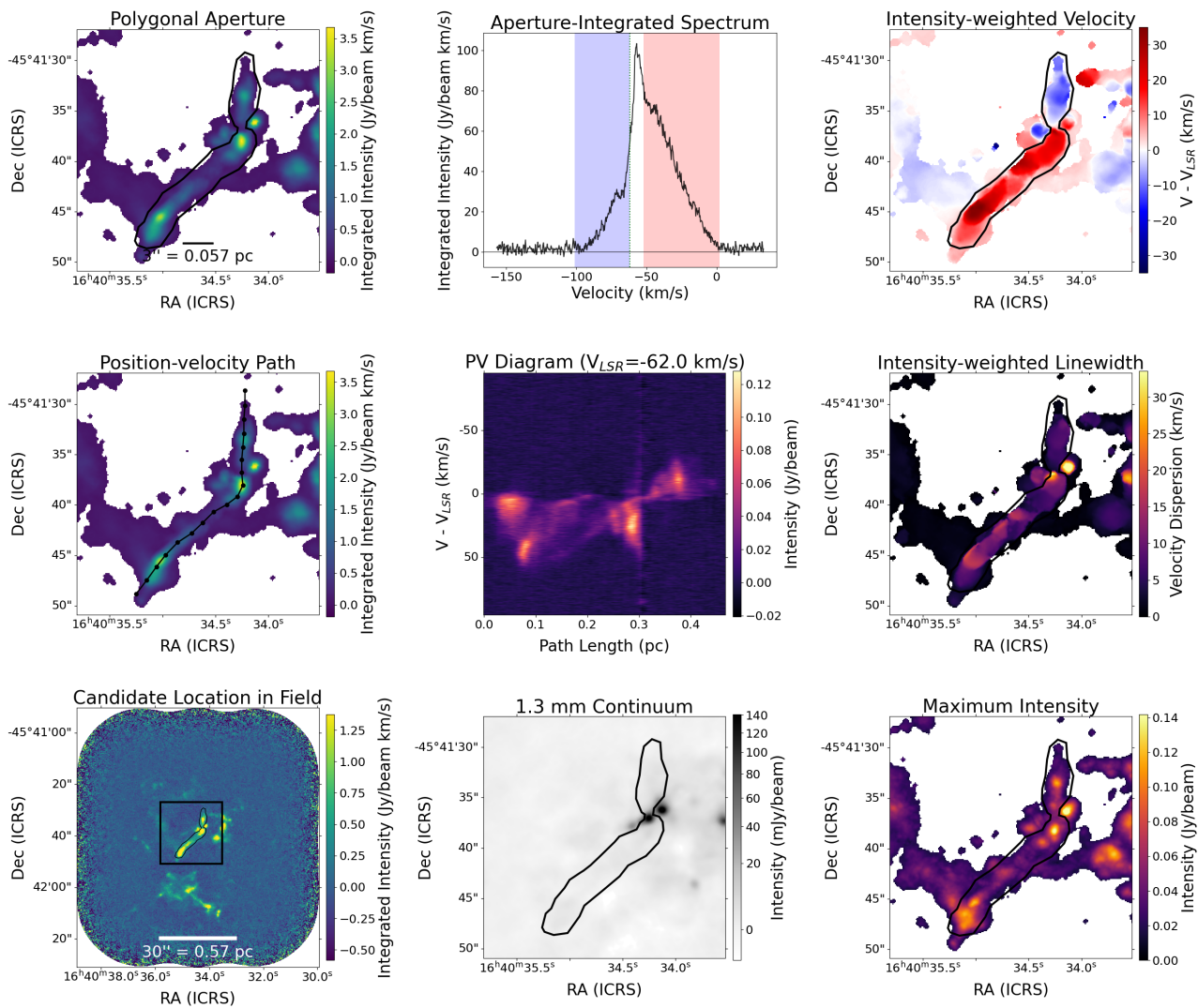

**Figure 2.** Same as in Figure 1, but for the asymmetric bipolar outflow Candidate #9 in field G338.93.

Outflow activity and outflow-core associations will be explored separately using a multi-species approach for certain individual targets in future publications (e.g. Armante et al., in prep for G012.80/W33 and Valeille-Manet et al., in prep for all massive cores in the 15 ALMA-IMF regions).

We stress that each outflow candidate remains an outflow *candidate*. We also note that many of our candidates do not have clear 1.3 mm continuum peaks within the outflow path or immediately adjacent to one end or the other of the longest axis, i.e. do not have candidate driving sources. In these cases, we suggest that either the SiO emission is tracing only the leading edge of the flow, or that this is an observational limitation driven by our sensitivity in each field. Both possibilities are consistent with the picture of protostellar outflows in which SiO preferentially traces the (smaller) active shocks and warm/hot gas in outflows and lower-J CO transitions trace the (larger) coolest, outer layers of an outflow's cavity walls (Bally 2016, Section 2).

The total number of bipolar candidates in our sample should be considered a lower limit. The low overall fraction of bipolar candidates is a result of the interactions between our identification method, classification criteria, and the highly-clustered nature of the dust-core populations in most fields. These limitations (single-species data, clustered dust cores) also affect our ability to positively identify continuum driving sources for a number of our candidates. There are many fields (especially but not only G012.80, G333.60, W43-MM1, W51-E, and W51-IRS2) in which numerous red- and blue-shifted outflows appear to be emanating from a single location in which two or more (usually >4) closely-packed dust cores are also located. In these cases, even slight misalignments in position angle between outflow candidates introduce significant uncertainty as to which candidates are associated with which cores, and thus with each other. We elected to leave most candidates in these confused regions classified as red or blue monopolar, rather than bipolar, because we are confident in their candidacy as outflows but less confident as to their specific red/blue pairings.

Some targets have overlapping fields of view (W43-MM2 & W43-MM3; W51-E & W51-IRS2). Consequently, four outflow candidates are detected in more than one field: W43-MM3 candidate #1 and candidate #2 are also detected at the eastern edge of the W43-MM2 field of view, and W51-IRS2 candidate #46 and candidate #47 are also detected in the northwestern quadrant of the W51-E field of view. We analyze these candidates using the W43-MM3 and W51-IRS2 data cubes, respectively, as these cubes had higher signal-to-noise ratios at the locations of these candidates. These candidates appear in the catalog under W43-MM3 and W51-IRS2 only, i.e. the entries are not duplicated under the alternate fields.

There were several findings in our SiO data set that are scientifically interesting, but beyond the scope of this catalog paper. These findings are briefly described in Appendix A, and will be explored in future publications.

### 3.3.1. Flux Filtering on Large Spatial Scales

Our 12-meter data were observed in either configuration C43-2, C43-3, or C43-1+4 combined. At 217.10498 GHz, the Maximum Recoverable Scale (MRS) of C43-1 (which sets the MRS of the C43-1+4 combined data) is $13''.1$, the MRS of C43-2 is $10''.4$, and the MRS of C43-3 is $7''.4$. The field of view (FOV) of a single pointing is $28''$ in all cases. In this section, we quantify the potential impact that complete or partial spatial filtering might have on the derived flux densities of our outflow candidates.

We tested the effect of spatial filtering by creating and imaging a synthetic "outflow" using CASA's `simobserve` and `simanalyze` tools. We created a FITS image with a 2-dimensional Gaussian with major axis of $16''$, minor axis $2''$, position angle of $-35°$, and integrated flux density of 3 Jy using `simobserve`. The flux density and major and minor axes are typical of the sizes and flux densities of our largest outflow candidates in their peak channel. The major and minor axes were also selected so that, in all three configurations, the "outflow" minor axis is resolved ($\geq 2 \times$ beam minor axis) and the major axis is larger than the MRS of the simulated observations. The position angle was chosen to avoid alignment with the pixel axes and the simulated beam major and minor axes.

**Table 3.** Catalog of SiO Outflow Candidates (Abbreviated)

| Field | ID | RA[a] | Dec[a] | Color[b] | $V_{range}^c$ | $V_{peak}^d$ | $F_{peak}^e$ | $F_{tot}^f$ | Classification |
|---|---|---|---|---|---|---|---|---|---|
| | # | $(^h\ ^m\ ^s)$ | $(^\circ\ '\ '')$ | | $(km\ s^{-1})$ | $(km\ s^{-1})$ | $(mJy)$ | $(Jy)$ | |
| G008.67 | 1 | 18:06:18.796 | -21:37:20.71 | red+blue | -19.0,30.0 | 28.94 | 790 (10) | 30.2 (0.2) | likely |
| | | | | | 40.0,119.5 | 45.11 | 290 (10) | 9.8 (0.2) | |
| G008.67 | 2 | 18:06:18.671 | -21:37:11.05 | red | 40.0,80.0 | 40.4 | 380 (10) | 22.0 (0.1) | likely |
| G008.67 | 3 | 18:06:19.734 | -21:37:19.81 | blue | -31.5,30.0 | 24.55 | 295 (9) | 19.1 (0.1) | likely |
| G008.67 | 4 | 18:06:19.071 | -21:37:23.53 | red+blue | 17.0,30.0 | 21.18 | 39 (2) | 0.74 (0.01) | possible |
| | | | | | 40.0,47.0 | 40.73 | 34 (2) | 0.297 (0.007) | |
| G008.67 | 5 | 18:06:19.093 | -21:37:27.85 | blue | 22.5,30.0 | 29.61 | 91 (3) | 1.10 (0.01) | possible |
| G008.67 | 6 | 18:06:23.589 | -21:37:04.33 | red | 43.0,65.0 | 48.15 | 377 (7) | 9.39 (0.06) | possible |
| G010.62 | 1 | 18:10:28.000 | -19:55:46.20 | blue | -20.0,-7.0 | -7.17 | 247 (1) | 3.504 (0.008) | possible |
| G010.62 | 2 | 18:10:28.160 | -19:55:47.19 | blue | -20.0,-9.0 | -8.85 | 549 (1) | 4.943 (0.008) | possible |
| G010.62 | 3 | 18:10:28.183 | -19:55:48.84 | blue | -31.0,-7.0 | -7.17 | 114 (1) | 2.53 (0.01) | possible |
| G010.62 | 4 | 18:10:28.269 | -19:55:36.58 | blue | -24.0,-11.0 | -14.92 | 217 (2) | 4.38 (0.01) | possible |
| G010.62 | 5 | 18:10:28.421 | -19:55:49.12 | red+blue | -38.0,-6.0 | -6.15 | 933 (3) | 23.75 (0.03) | complex or cluster |
| | | | | | 4.0,20.5 | 3.96 | 530 (3) | 7.33 (0.02) | |
| G010.62 | 6 | 18:10:28.655 | -19:55:49.83 | red+blue | -20.0,-7.5 | -7.5 | 227 (3) | 1.72 (0.02) | possible |
| | | | | | 2.5,40.0 | 2.61 | 472 (3) | 8.35 (0.03) | |

[a]The ICRS coordinates of the center of the bounding box for each polygonal region. Uncertainties on each position are ±1 pixel, where pixel sizes are listed in Table 2.

[b]Bipolar outflow candidates (classified as "red+blue") have their properties listed on two lines instead of a single line; the first line is always the blue lobe and the second line is always the red lobe. Red and blue lobes in a bipolar candidate have the same overall classification, i.e. both "likely," "possible," or "complex or cluster."

[c]Velocity range is identified from aperture-integrated intensities and position-velocity diagrams and shown in the upper-right panels of Figures 1 and 2. In most cases, the velocity range $-5\ km\ s^{-1} < V_{LSR,candidate} < +5\ km\ s^{-1}$ is excluded. In some rare cases, we exclude more or less of the velocity range around $V_{LSR,candidate}$, based on line shape in the integrated spectrum.

[d]The velocity at which the aperture-integrated spectrum peaks, *within the velocity range listed in the preceding column.* In other words, this is the peak within an outflow candidate excluding the ambient emission.

[e]The peak of the aperture-integrated flux density.

[f]The total aperture- and velocity-integrated flux density of the candidate.

We simulated observations of this Gaussian with the *uv*-coverage and typical integration times for our data, and generated simulated measurement sets (MS). We imaged these MS files interactively using multiscale deconvolution in `tclean`. We cleaned all three images to 5 mJy, with cell sizes 1/5 the size of the beam minor axis in each configuration, and created primary beam-corrected versions of each image. After imaging, we drew a single polygonal aperture that encompassed the flux of our "outflow" in all three cleaned images. We compared the aperture-integrated flux density of each image to the aperture-integrated flux density of our simulated Gaussian component.

We find that the effect of spatial filtering on our measured flux is always <5%. Surprisingly, we found that the filtered, cleaned data overestimated the flux by 2-3% (C43-2, C43-1+4) or 5% (C43-3). We attribute this excess to flux being pushed into the sidelobes and then included in the measurement aperture - essentially a consequence of having a finite measurement aperture but a non-finite extent to the Gaussian model.

We conclude that any effect of spatial filtering on our measured flux densities is small, and adopt a flux-density uncertainty of 5% for all data sets.

### 3.3.2. *Comparison to Nony et al. (2023)*

In order to evaluate potential biases in our outflow identification methodology, we compare the outflows we identify in W43-MM2 and W43-MM3 with those identified by Nony et al. (2023) for these same fields. Nony et al. (2023) use both $^{12}$CO (2-1) and SiO (5-4) line cubes to identify protostellar outflows by eye in W43-MM2 and W43-MM3. They require emission of $\geq 5\sigma$ in 3 consecutive channels in the CO cube only in order to identify a candidate. Nony et al. (2023) center their search on dust cores specifically, as their goal is to distinguish prestellar from protostellar cores through the presence of outflow emission.

We find that results agree well with those of Nony et al. (2023). Using our polygonal apertures and those of Nony et al. (2023), we integrate the flux density of each candidate from $-95$ km s$^{-1} \leq \mathrm{V}_{LSR} \leq +95$ km s$^{-1}$, excluding the central 10 km s$^{-1}$ ($\mathrm{V}_{LSR} \pm 5$ km s$^{-1}$). The two methods are largely similar when it comes to bright candidates, with maximum flux densities from the Nony et al. (2023) apertures of 53 Jy and 20 Jy for W43-MM2 and W43-MM3, respectively, compared to our maxima of 51 Jy and 19 Jy, respectively. Overall, the apertures of Nony et al. (2023) capture fainter emission, with minimum flux densities of 0.09 Jy and 0.06 Jy for W43-MM2 and W43-MM3, respectively, compared to the minimum flux densities of 0.4 Jy and 0.5 Jy, respectively, obtained with the apertures used in this work. The total outflow flux densities within each region (sum of the fluxes of each individual candidate) agree within 1% for W43-MM2 (197 Jy versus 195 Jy) and 15% for W43-MM3 (67 Jy versus 57 Jy) between the Nony et al. (2023) apertures and our apertures, respectively. As the emission in W43-MM3 overall skews fainter than that in W43-MM2, this larger deviation in results for W43-MM3 is expected.

We find more variation between our results and those of Nony et al. (2023) when we consider each candidate individually. In W43-MM2, we identify 27 outflow candidates in SiO, while Nony et al. (2023) identify 33 candidates in SiO+CO and 14 candidates in SiO only. In W43-MM3, we identify 13 outflow candidates in SiO, while Nony et al. (2023) identify 14 candidates in SiO+CO and 1 candidate in SiO only. The most significant differences between our identifications are in outflow morphology for outflows in common between the two methods, and in the detection/non-detection of candidates with very low signal-to-noise ratios ($\lesssim 3\sigma$). There are also several cases in which features we identify as being separate candidates in our SiO data are identified as single, larger outflows in Nony et al. (2023) using CO data. Of the candidates identified by Nony et al. (2023) that are not identified in our catalog (21 candidates in W43-MM2, 5 in W43-MM3), all are either detected by Nony et al. (2023) in both SiO and CO − which allows those authors to probe fainter, less continuous SiO emission with greater confidence − or are de-tected in areas of our maps that are completely masked. In other words, Nony et al. (2023) include weaker SiO emission in their identifications than we do, and this accounts for the difference in the total numbers of identified outflows. The candidates identified by us that are not identified by Nony et al. (2023) (3 candidates in W43-MM2, 5 in W43-MM3) are all both a) smaller and less elongated than the majority of our sample, and b) lacking in any obvious 1.3 mm dust core candidates in or near the outflow apertures. This difference is to be expected, as Nony et al. (2023) are specifically searching for outflow emission around dust cores, while we do not limit our search in this way.

Overall, we find that our SiO-only, spatially-unbiased search method is less sensitive to faint or spatially-incoherent outflow emission than a dust core-centered, CO-based search method. However, we find that the field-aggregated impact of this sensitivity bias is minimal, with total methodological uncertainties of ∼15% at most. We also find that our method captures some emission missed by dust-core-centered search methods. Because we limit ourselves to discussing field-aggregated outflow properties in § 4, rather than individual candidates, any candidate-level biases are unlikely to have a significant impact on the results presented in this work.

### 3.3.3. Crossing Outflows

There are several outflow candidates in our catalog which overlap with each other both spatially (along the line of sight) and spectrally (with overlapping velocity ranges). In these cases, the emission from one outflow candidate effectively "contaminates" the other. In order to account for this contamination, we identify the area of spatial and spectral overlap from the two outflows and then calculate the total flux density in this overlap region. If all of the overlap-region flux truly belonged to Outflow A, this would reduce the flux density of Outflow B, and vice-versa. In other words, the overlap-region flux is essentially a maximum possible contamination level. We therefore add this overlap flux to the lower-bound uncertainty of each of the crossing outflows. These entries have asymmetric uncertainties in the complete version of Table 3. In most cases, the flux density of the overlap region is ≤15% of the total flux density of each candidate. The overlap flux density is only ≳33% in two cases: G012.80 Candidate #30 (red lobe only, 68%) and W51-E Candidate #18 (blue lobe only, 48%).

This method does have the effect of "double-counting" the flux in the overlap region, because we do not subtract it from either candidate. When we discuss the field-aggregated outflow properties (§ 4), this will have the effect of increasing the lower-bound uncertainties on the field-aggregated values. However, we find that this impact is minimal, as the field-total uncertainties remain small relative to the field totals regardless of how many crossing outflows a field contains.

### 3.3.4. Outflow Inclination Angle

The observed velocity of each outflow only captures the line-of-sight component of the true velocity vector. Likewise, plane-of-the-sky projection effects mean that our derived outflow lengths are lower limits. This affects the derived properties that depend on velocity or outflow length, i.e. all properties except mass.

We do not attempt to measure unique outflow inclination angles for our candidates. Therefore, we report outflow properties for each candidate without any inclination correction in this paper and in our online tables. Inclination-corrected representative statistics for the full sample only are presented in § 3.4.

### 3.4. Physical Properties of the Outflow Candidates

For each outflow candidate, we derive its median SiO column density and its total mass, momentum, kinetic energy, mass rate, momentum rate, and energy rate. These properties are presented for each candidate in a machine-readable table in the online material; a representative example is shown in Table C1.

We first convert each cube from Jy beam$^{-1}$ to K, and then extract the spatially-integrated spectrum of each candidate using the custom polygonal apertures described in § 3.2. These aperture-integrated spectra are the basis of our derivation of the physical properties of each candidate. We use a channel-based calculation method (see, e.g. Maud et al. 2015) in which each physical property is calculated separately for each channel and then summed, rather than using velocity-integrated fluxes. This channel-based method has been shown to reduce the overestimation of outflow momenta and kinetic energies that can occur when aperture-integrated intensities are multiplied by an outflow's maximum velocity only (see Maud et al. 2015, Section 3.3 and references therein).

### 3.4.1. Derivation of Column Density

After converting our cube to units of Kelvin and extracting the aperture-integrated spectrum for a candidate, we mask out any channels in which the aperture-integrated brightness temperature is <3σ, where σ is the aperture-integrated noise level. This masking step helps to prevent high-velocity, low-signal features from disproportionately impacting both the derived momenta and energies and their associated uncertainties at later stages.

To derive column density in each channel individually, we adopt a discrete form of the general equation for molecular column density in the optically thin approximation (see Appendix B):

$$N(i) = \frac{8\pi k_B \nu^2}{hc^3 A_{ul}} \frac{Q_{rot}}{g_J g_K g_I} \exp\left(\frac{E_u}{k_B T_{ex}}\right) T_{B,i} \Delta v \quad (1)$$

where $k_B$ is the Boltzmann constant, $\nu$ is the rest frequency of the SiO $J = 5-4$ transition, $h$ is the Planck constant, $c$ is the speed of light, $A_{ul}$ is the Einstein coefficient of spontaneous emission from the upper state to the lower state, $Q_{rot}$ is the partition function of the SiO molecule, $g_J$, $g_K$, and $g_I$ together represent the total degeneracy of the rotational state, $E_u$ is the energy of the upper state of the transition, $T_{ex}$ is the excitation temperature, $T_{B,i}$ is the aperture-integrated brightness temperature in channel $i$, and $\Delta v$ is the channel width. $N(i)$ is the aperture-integrated column density in channel $i$. We assume that the SiO emission is optically thin, and assume an excitation temperature of T=$50^{+30}_{-20}$ K for all candidates. A detailed discussion of our reasons for these choices can be found in Appendix B.

The non-linear relationship between $T_{ex}$ and $N_{SiO}$ leads to asymmetric uncertainties in our column densities, and in all properties that are subsequently derived from them (mass, momentum, energy, and associated rates). We propagate this asymmetric uncertainty by calculating two Gaussian uncertainties for each $N(i)$: one using $\sigma_{T_{ex,lower}} = -20$ K, and one using $\sigma_{T_{ex,upper}} = +30$ K. The total error distribution for each derived $N(i)$ becomes the combination of two half-Gaussians: below the calculated $N(i)$ value, it is a Gaussian with $\sigma = \sigma_{lower}$ and centered at $N(i)$, and above the calculated $N(i)$, it is a Gaussian with $\sigma = \sigma_{upper}$ centered at $N(i)$. When summing the data either spatially or spectrally, the lower- and upper-bound uncertainties are summed in quadrature separately, i.e.

$$\sigma_{N\,lower,\,sum} = \left(\sum_i (\sigma_{N\,lower,i})^2\right)^{1/2} \quad (2)$$

and likewise for $\sigma_{upper}$. We find that our column density uncertainties can be dominated either by $\sigma_{T_{ex}}$ or by the inherent noise in the data itself, depending on the data noise level. Fields with higher median $\sigma$ (see Table 2) tend to be noise-dominated, and those with lower median $\sigma$ tend to be dominated by the uncertainty in $T_{ex}$.

### 3.4.2. Derived Masses, Momenta, and Kinetic Energies

The total gas mass in each channel can be calculated following

$$M_{gas}(i) = N(i)\mu_g m_{H_2}\Omega D^2 / \left[\frac{SiO}{H_2}\right] \quad (3)$$

where $\mu_g = 1.36$ is the total gas mass relative to $H_2$. $m_{H_2}$ is the mass of a hydrogen molecule, $\Omega$ is the solid angle subtended by a single pixel in our image cubes, $D$ is the distance to the source, N(i) is given by Equation 1, $\left[\frac{SiO}{H_2}\right]$ is the fractional abundance of SiO relative to molecular hydrogen.

We adopt a flat SiO-to-$H_2$ abundance ratio of $10^{-8.5}$ (or, $3.16 \times 10^{-9}$) for all outflow candidates in our sample, taking into consideration the wide range of abundance ratios in astrochemical shock models (Schilke et al. 1997; Gusdorf et al. 2008), intra-outflow abundance variations (Bally 2016), and typical abundance ratio values reported in the literature (see e.g. Codella et al. 1999; Lu et al. 2021). A detailed discussion of our choice of $\left[\frac{SiO}{H_2}\right]$ can be found in Appendix B. In short, these theoretical and observational studies have shown abundance ratios ranging from $10^{-11}$ to $10^{-6}$ both within individual outflows and sometimes between outflows. Such large variations are not guaranteed to occur within any given outflow, but they are possible across a population. This means that our adopted ratio of $10^{-8.5}$ could potentially vary by up to two orders of magnitude from the true abundance at any single location within an individual outflow candidate.

However, because we are unable to measure the SiO abundance ratio directly (see Appendix B, Section 3), we also do not have a measurement uncertainty for our assumed abundance ratio. For the purposes of error propagation, we therefore treat the fractional abundance as definitional (i.e. $\sigma = 0$). This assumed fractional abundance may be an overestimate for some sources and an underestimate for others. This would increase the scatter in our data at the population level, but it is unlikely to change underlying fundamental correlations or distributions unless there is a trend in this over- or under-estimation. We compare our data to the SiO-derived outflow masses in the literature (including Lu et al. 2021) in § 4.1 and find no evidence for such a trend.

With the adoption of $\left[\frac{SiO}{H_2}\right] = 10^{-8.5}$, Equation 3 becomes

$$M_{gas}(i) = \left(1 \times 10^{8.5}\right) N(i)\mu_g m_{H_2}\Omega D^2 \qquad (4)$$

alternately, substituting Eq. 1 for $N(i)$ gives us

$$M_{gas}(i) = C\sum_i T_B(i) \qquad (5)$$

where $T_B(i)$ is the aperture-integrated brightness temperature of a single channel and

$$C = 2.08 \times 10^{-8}\Delta v \Omega D^2 \qquad (6)$$

is the constant of proportionality between brightness temperature and mass for SiO. In Equation 6, the leading numerical factor has units of [g s cm$^{-3}$ K$^{-1}$], $\Delta$v is in cm s$^{-1}$, $\Omega$ is in steradians, and D is in cm.

The total, velocity- and spatially-integrated mass is then

$$M_{out} = \sum_i M_{gas}(i) \qquad (7)$$

where the summation is performed over all channels in the outflow's unique velocity range (see the complete online version of Table 3 for velocity ranges).

We then calculate outflow momentum as

$$P_{out} = \sum_i M(i)v_i \qquad (8)$$

and outflow kinetic energy as

$$E_{out} = \frac{1}{2}\sum_i M(i)v_i^2 \qquad (9)$$

where $v_i$ is the velocity of channel $i$ relative to local $V_{LSR}$ in both Eqs. 8 and 9. As discussed in § 3.3.4, we do not assume an inclination angle when deriving these properties.

Figure 3 shows the log-space distribution of these properties for the full sample (all 315 candidates, with the red and blue lobes of bipolar candidates counted separately for a total of 355 data points in the plotted bins). All three panels show stacked histograms.

In general, the distributions have well-defined peaks but are broad, spanning >3.5 orders of magnitude in mass, >4 orders of magnitude in momentum, and >5 orders of magnitude in kinetic energy. Minimum, maximum, and median values for mass, momentum, and energy for the full sample are listed in the upper section of Table 4. These values are not adjusted for inclination angle.

We derive inclination-adjusted mass, momentum, and energy statistics for the full sample assuming a uniform inclination angle of $\sim$57.°3, following the method of Bontemps et al. (1996). These statistics are listed in the lower section of Table 4, as are the inclination correction factors for each property. The inclination-corrected statistics are not used in our analysis unless specifically noted.

Table C1 shows our derived median column density, mass, momentum, and kinetic energy for the red- and blue-shifted components of each outflow candidate. The first ten lines are shown in the print version of this paper. The full, machine-readable version of this table, including uncertainties for each value, can be viewed

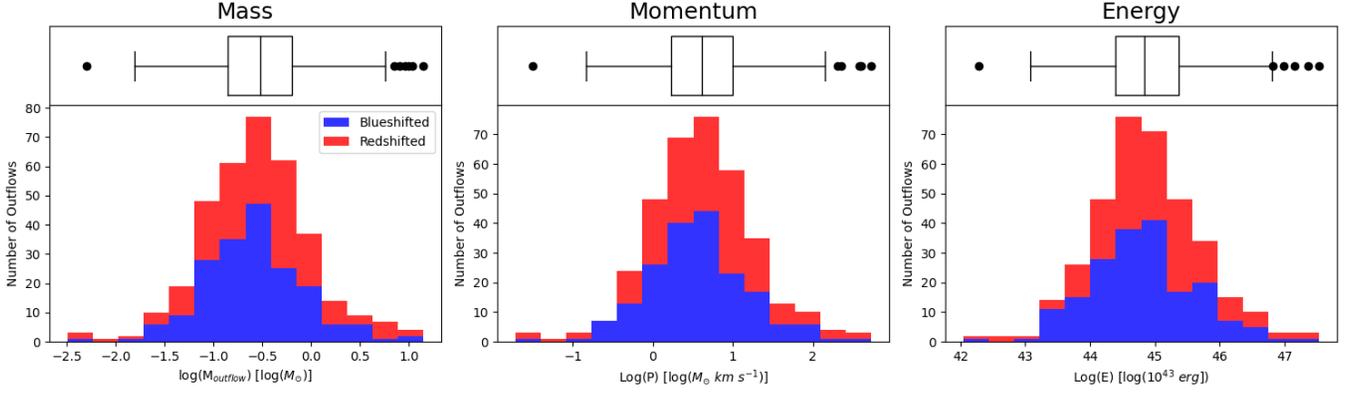

**Figure 3.** The distribution of mass, momentum, and kinetic energy for all 315 outflow candidates. Red bars indicate redshifted outflows, and blue bars indicate blueshifted outflows. The histogram is stacked. The box-and-whisker plots above each histogram are for the total (red and blue combined) outflow population. The central line in each box indicates the median value, the left and right edges of the box indicate the first and third quartiles, respectively, the left and right whiskers extend from the first and third quartiles by $1.5\times$ the inter-quartile range, respectively, and outlier ("flier") points are represented by black circles.

**Table 4.** Full Sample Mass, Momentum, & Energy Statistics

| Property | Min | Max | Median[a] |
|---|---|---|---|
| $M_{blue}$ [M$_\odot$] | 0.003 | 10.0 | 0.25 (0.15) |
| $M_{red}$ [M$_\odot$] | 0.003 | 14.1 | 0.34 (0.23) |
| $M_{tot}$ [M$_\odot$] | 0.005 | 14.1 | 0.30 (0.20) |
| $P_{blue}$ [M$_\odot$ $km$ $s^{-1}$] | 0.019 | 402 | 3.4 (2.4) |
| $P_{red}$ [M$_\odot$ $km$ $s^{-1}$] | 0.030 | 538 | 4.7 (3.5) |
| $P_{tot}$ [M$_\odot$ $km$ $s^{-1}$] | 0.031 | 538 | 4.1 (2.9) |
| $E_{blue}$ [$erg$] | $1.1\times10^{42}$ | $2.3\times10^{47}$ | $5.8\times10^{44}$ (4.8) |
| $E_{red}$ [$erg$] | $1.9\times10^{42}$ | $3.4\times10^{47}$ | $8.3\times10^{44}$ (7.1) |
| $E_{tot}$ [$erg$] | $1.9\times10^{42}$ | $3.4\times10^{47}$ | $6.8\times10^{44}$ (5.6) |
| | | | |
| Inclination-adjusted ($i = 57.3°$) | | | |
| | | | |
| $P_{blue}$ / $\cos i$ | 0.035 | 744 | 6.2 (4.3) |
| $P_{red}$ / $\cos i$ | 0.056 | 996 | 8.7 (6.5) |
| $P_{tot}$ / $\cos i$ | 0.057 | 996 | 7.6 (5.3) |
| $E_{blue}$ / $\cos^2 i$ | $3.8\times10^{42}$ | $8.0\times10^{47}$ | $2.0\times10^{45}$ (1.6) |
| $E_{red}$ / $\cos^2 i$ | $6.5\times10^{42}$ | $1.2\times10^{48}$ | $2.8\times10^{45}$ (2.4) |
| $E_{tot}$ / $\cos^2 i$ | $6.5\times10^{42}$ | $1.2\times10^{48}$ | $2.3\times10^{45}$ (1.9) |

[a] The min, max, and median values reported in this table are calculated from the full sample of 315 candidates.

[b] Uncertainties on the medians are the scaled MAD, listed in parentheses. For values listed in scientific notation, the order of magnitude of the uncertainty is the same as that of the reported median.

[c] The values in the upper section of the table are not adjusted for inclination angle. The values in the lower section assume a uniform inclination angle of $57.°3$ for all candidates.

in the online Journal and in ESCV format on Zenodo: 

Overall, we find no statistically-significant difference between mass, momentum, or energy values for the red versus blue outflow lobes; this is consistent with a lack of any strong detection bias toward strong or weak emission with lobe color. Our high energy maxima can be attributed to those outflows which have both bright emission at all velocities and strong emission at $|$V $-$ V$_{LSR}|$ $\geq$60 km s$^{-1}$. Because energy goes as v$^2$, emission at high velocities has an outsized effect on the total derived energy. In most cases this high-velocity gas is all part of the outflow, but in a small subset of cases this "high velocity" emission is due to hot-core line emission contaminating the outflow aperture. This appears as "high-velocity" emission because the lines are at different rest frequencies from SiO 5-4, and so this hot-core line contamination has a strong effect on the derived energies in these few cases. There are 12 outflow candidates within the sample with significant contamination from hot-core lines: G351.77 Candidate #3, W43-MM1 Candidates #16, #17, and #27, W43-MM2 Candidates #14 and #15, W51-E Candidates #19 and #20, and W51-IRS2 Candidates #10, #28, #38, and #40. The derived properties of these candidates should therefore be considered upper limits. The hot cores in each field are explored separately in Brouillet et al. (2022) and Bonfand et al. (2023, submitted).

### 3.4.3. *Derived Dynamical Times and Mass, Momentum, and Energy Rates*

In order to determine mass flow rate $\dot{M}$, momentum supply rate $\dot{P}$ (alternately mechanical force, $F_m$), and outflow power $\dot{E}$ (alternately mechanical luminos-ity $L_m$), most teams measure the distance between an outflow and its driving source and divide this by outflow velocity in order to determine a rough outflow dynamical time. This approach requires the identification of a continuum driving source for each candidate. Since we do not (and in many cases cannot) assign our candidates to specific driving sources in this work, we cannot use this approach. Instead, we use the position-velocity path length (see Figures 1 and 2) as a proxy for outflow size. Our position-velocity path length is the same in all channels, so a channel-by-channel calculation of outflow dynamical time is not possible with this approach. Instead, we divide the outflow path length by the median relative velocity (median intensity-weighted velocity minus $V_{LSR}$) to determine a single $t_{dyn}$ for each candidate.

We assign an uncertainty of $\pm 15\%$ to all outflow dynamical times. This uncertainty is a consequence of our use of path length as a proxy for outflow size. When creating the position-velocity paths, we sometimes extend the path beyond the end of the outflow in order to include baseline zero-emission regions in the PV diagram. Likewise, in very crowded regions, path lengths are truncated slightly to avoid confusion with nearby features. In all cases, the difference between the path marked in CARTA as the "true" outflow size (identified by eye in the moment maps) is $\lesssim 15\%$. Therefore, this is the uncertainty we adopt for the path length and outflow dynamical times.

Our median dynamical time from this method is 6,000 $\pm$ 2,800 years (where the uncertainty is the scaled MAD), and our minimum and maximum dynamical times are 510 years and 70,000 years, respectively. Outflow mass rate is then $M_{out}/t_{dyn}$, outflow momentum rate is $P_{out}/t_{dyn}$, and outflow energy rate is $E_{out}/t_{dyn}$.

We do not calculate a dynamical time for any candidates classified as "complex or cluster," as for these candidates the path is arbitrary and does not reflect a specific outflow axis. This criterion eliminates 6 candidates from further analysis. We also do not calculate dynamical time or associated rates for any candidates whose path length is less than twice the length of the beam major axis, i.e., candidates whose largest axis remains unresolved. Typically, these are candidates suspected of having a face-on orientation. This criterion eliminates 5 more candidates from further analysis. In total, we reduce our total number of candidates to 304 for the analysis of mass, momentum, and energy rates and all associated figures.

Figure 4 shows the log-space distribution of these rates for the full sample. All three panels show stacked histograms. Table 5 shows the minimum, maximum, and median values for each derived rate. The values in the upper section of the Table 5 are not adjusted for inclination angle. Inclination-adjusted values are listed in the lower section of Table 5, as are the inclination correction factors for each property. We assume a uniform inclination angle of $57.^\circ$ for all candidates to derive these values. The inclination-corrected values are not used in our analysis unless specifically noted.

As in § 3.4.2, we find no significant differences between the rates derived for blueshifted outflow lobes and those derived for redshifted lobes. Likewise, we find that each rate has a reasonably well-defined peak but that the distributions are again broad, spanning 3.7-4.2 orders of magnitude in mass rate, 4.6-5 orders of magnitude in momentum rate, and 5.5-6 orders of magnitude in kinetic energy rate.

## 4. DISCUSSION

### 4.1. Comparison with Similar Samples

In this section, we compare the physical properties derived for our sample to those derived in the literature for similar high-mass samples. Specifically, we compare to lower-resolution, single-dish SiO (Csengeri et al. 2016) and CO (Maud et al. 2015) observations of larger massive protocluster samples than ours, and to a smaller sample of protoclusters detected in SiO with similar spatial resolution to our data (Lu et al. 2021). This comparison allows us to test both methodological effects and the impact of interferometric versus single-dish data, as well as placing our findings in broader context with the literature. We find good agreement between our typical (median, minimum/maximum) outflow derived properties and those reported in the literature for these other samples, suggesting that any methodological or observational bias in our candidate identification procedure or derivation of physical properties has not had a significant impact on our results. We describe our comparison to each sample in greater detail in the following paragraphs.

We first compare our derived outflow column densities to those derived by Csengeri et al. (2016) for a sample of massive clumps selected from the ATLASGAL survey. Csengeri et al. (2016) observe 430 sources with the IRAM 30-meter telescope at 84~115 GHz ($\sim 26''$ beam), and a subsample of 128 sources with the APEX telescope at 217 GHz ($29''$ beam). For their full sample, Csengeri et al. (2016) derive column densities of $1.6 \times 10^{12} - 7.9 \times 10^{13}$ cm$^{-2}$ using SiO J= 2−1 data and assuming an LTE approximation. For their subsample of 128 sources measured with both SiO J=2−1 and SiO J=5−4, Csengeri et al. (2016) find a column density range of $9.6 \times 10^{11} - 1.4 \times 10^{14}$ cm$^{-2}$ using RADEX mod-eling, depending on the treatment of the beam filling factor. Our median column densities for any individual field range from $5.0\times10^{13}$ cm$^{-2}$ to $3.5\times10^{14}$ cm$^{-2}$, and the median column density across all 315 candidates ranges from $9\times10^{12}$ to $1.2\times10^{15}$ cm$^{-2}$, with a median of $1\times10^{14}$ cm$^{-2}$. These values are reported in Table 6. Our SiO column densities overlap with those of these ATLASGAL sample, but trend $\sim$1 order of magnitude higher overall. This trend is likely due to differences in angular resolution, as the Csengeri et al. (2016) single-dish data are more strongly affected by beam dilution than our interferometric data.

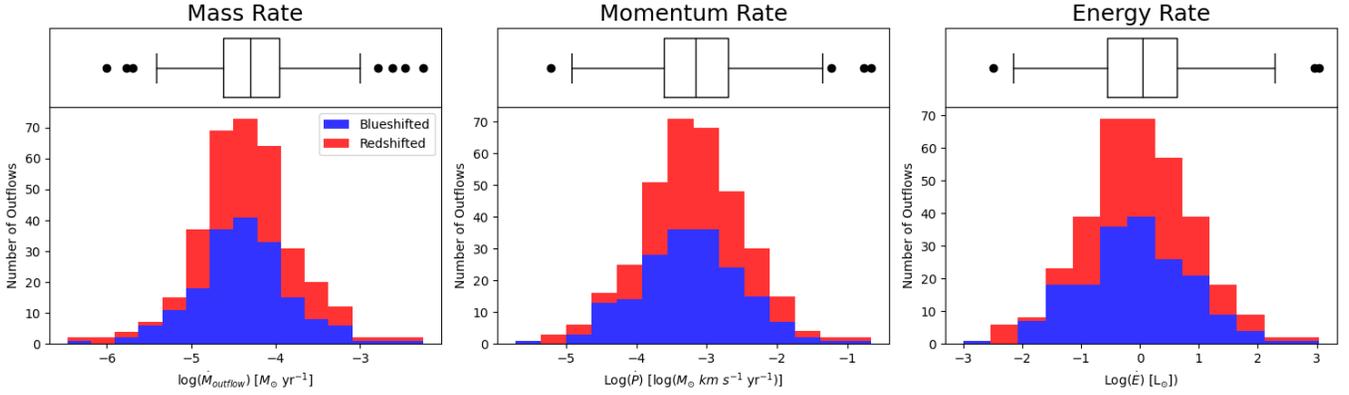

**Figure 4.** The distribution of mass rate, momentum rate, and energy rate for all 304 outflow candidates (315 candidates minus the 11 candidates either classified as "complex or cluster" or unresolved along their longest axis). Red bars indicate redshifted outflows, and blue bars indicate blueshifted outflows. The histogram is stacked. Box-and-whisker plots have the same meaning as in Figure 3, but for the slightly smaller population of 304 candidates. The highest values in the energy-rate histogram should be taken as upper limits due to contamination from hot-core line emission.

We compare our derived outflow masses to those derived by Lu et al. (2021) for their sample of massive star-forming regions in the Central Molecular Zone (CMZ). Lu et al. (2021) target 834 molecular clumps with 2000 au resolution and detect 43 outflows. They derive a separate mass for each outflow using each of their molecular tracers; their SiO-derived outflow masses range from a few hundredths to a few tens of solar masses. The masses we derive for our outflow candidates largely fall within this range (see Table 4), but our minimum derived masses extend $\sim$1 order of magnitude lower than those of Lu et al. (2021). This can likely be attributed to the greater distance to the CMZ ($>$8 kpc), which will result in decreased mass sensitivity. This consistency is notable considering Lu et al. (2021) derive position-dependent SiO abundance ratios for each outflow. The overall agreement between our mass range and theirs suggests that our choice of $\left[\frac{SiO}{H_2}\right] = 10^{-8.5}$ is a reasonable first-order approximation of SiO abundance at the population level. Though specific abundances may vary within or between individual outflows, the true average value in our data appears to be well-represented by $10^{-8.5}$ to first order.

Lu et al. (2021) find typical outflow velocities of several tens of km s$^{-1}$ and an overall range of a few km s$^{-1}$ to $>$90km s$^{-1}$ for their CMZ sample (see Lu et al. 2021, Table 3), comparable to those we derive for our candidates (see Table 3 and the complete online version).

We also compare our outflow properties to those of Maud et al. (2015), who use the JCMT to examine $^{12}$CO and $^{13}$CO J $=$ 3$-$2 emission toward 99 massive young stellar objects drawn from the Red MSX Source survey (RMS). For each outflow, Maud et al. (2015) derive its mass, momentum, kinetic energy, dynamical time, and mass, momentum, and energy rates. As theirs are single-dish data, Maud et al. (2015) only report a single red and blue outflow for each field.

In order to compare our results to Maud et al. (2015), we create "field-aggregated" outflow properties for each protocluster by summing each property (mass, momentum, etc) across all outflows in each field. These values are reported in Table 6. We find a median field-aggregated outflow mass of $8.74^{+0.09}_{-0.1}$ M$_\odot$, with a minimum of $1.40^{+0.04}_{-0.04}$ and a maximum of $56.6^{+0.9}_{-0.4}$. We find median, minimum, and maximum field-aggregated outflow momenta of $141^{+1}_{-1}$ M$_\odot$ km s$^{-1}$, $31.8^{+0.8}_{-0.9}$ M$_\odot$ km s$^{-1}$, and $1550^{+20}_{-10}$ M$_\odot$ km s$^{-1}$, and median, minimum, and maximum field-aggregated kinetic energies of $4.26^{+0.05}_{-0.06} \times 10^{46}$ erg, $0.9^{+0.01}_{-0.01} \times 10^{46}$ erg, and $73.3^{+0.9}_{-1.1} \times 10^{46}$ erg, respectively.

**Table 5.** Full Sample Mass, Momentum, & Energy Rate Statistics

| Property[a] | Min | Max | Median[b] |
|---|---|---|---|
| $t_{dyn}$ [yr] | 510 | 70000 | 6000 (2800) |
| $\dot{M}_{blue}$ [$M_\odot$ yr$^{-1}$] | $3.4\times10^{-7}$ | 0.006 | $3.9\times10^{-5}$ (2.7) |
| $\dot{M}_{red}$ [$M_\odot$ yr$^{-1}$] | $6.0\times10^{-7}$ | 0.003 | $4.7\times10^{-5}$ (3.3) |
| $\dot{M}_{tot}$ [$M_\odot$ yr$^{-1}$] | $1.0\times10^{-6}$ | 0.006 | $7.0\times10^{-5}$ (3.2) |
| $\dot{P}_{blue}$ [$M_\odot$ km s$^{-1}$ yr$^{-1}$] | $1.9\times10^{-6}$ | 0.22 | $6.1\times10^{-4}$ (4.9) |
| $\dot{P}_{red}$ [$M_\odot$ km s$^{-1}$ yr$^{-1}$] | $6.0\times10^{-6}$ | 0.17 | $8.0\times10^{-4}$ (6.1) |
| $\dot{P}_{tot}$ [$M_\odot$ km s$^{-1}$ yr$^{-1}$] | $6.0\times10^{-6}$ | 0.22 | $7.0\times10^{-4}$ (5.7) |
| $\dot{E}_{blue}$ [$L_\odot$] | 0.001 | 1100 | 0.9 (0.8) |
| $\dot{E}_{red}$ [$L_\odot$] | 0.003 | 900 | 1.1 (1.0) |
| $\dot{E}_{tot}$ [$L_\odot$] | 0.003 | 1100 | 1.1 (1.0) |
| **Inclination-adjusted[c] ($i = 57.3°$)** | | | |
| $t_{dyn}$ / $\tan i$ | 330 | 45000 | 3900 (1800) |
| $\dot{M}_{blue}$ ($\tan i$) | $5.3\times10^{-7}$ | 0.009 | $6.1\times10^{-5}$ (4.2) |
| $\dot{M}_{red}$ ($\tan i$) | $9.3\times10^{-7}$ | 0.005 | $7.3\times10^{-5}$ (5.1) |
| $\dot{M}_{tot}$ ($\tan i$) | $1.6\times10^{-6}$ | 0.009 | $7.8\times10^{-5}$ (5.0) |
| $\dot{P}_{blue}$ ($\sin i$ / $\cos^2 i$) | $5.5\times10^{-6}$ | 0.63 | 0.0018 (0.0014) |
| $\dot{P}_{red}$ ($\sin i$ / $\cos^2 i$) | $1.7\times10^{-5}$ | 0.49 | 0.0023 (0.0018) |
| $\dot{P}_{tot}$ ($\sin i$ / $\cos^2 i$) | $1.7\times10^{-5}$ | 0.55 | 0.0020 (0.0016) |
| $\dot{E}_{blue}$ ($\sin i$ / $\cos^3 i$) | 0.005 | 5900 | 4.8 (4.3) |
| $\dot{E}_{red}$ ($\sin i$ / $\cos^3 i$) | 0.018 | 4800 | 5.9 (5.3) |
| $\dot{E}_{tot}$ ($\sin i$ / $\cos^3 i$) | 0.018 | 5900 | 5.9 (5.3) |

[a] The min, max, and median values reported in this table are calculated from the full sample of 315 candidates.

[b] Uncertainties on the medians are the scaled MAD, listed in parentheses. For values listed in scientific notation, the order of magnitude of the uncertainty is the same as that of the reported median.

[c] The values in the upper section of the table are not adjusted for inclination angle. The values in the lower section assume a uniform inclination angle of $57.3°$ for all candidates.

We find that our field-aggregated outflow properties typically fall within the ranges observed by Maud et al. (2015), who find outflow masses of $\sim 0.7$ $M_\odot$ $- 1000$ $M_\odot$, outflow momenta of $\sim 3$ $M_\odot$ km s$^{-1}$ $- 4000$ $M_\odot$ km s$^{-1}$, and outflow kinetic energies of $\sim 10^{44}$ erg $- 3\times10^{47}$ erg for their outflows. We do note that our derived total outflow masses tend toward the lower end of the distribution observed by Maud et al. (2015) for their sample, our momenta are largely in line with the RMS-derived distribution, and our energies tend toward the higher end of the RMS-derived distribution. These trends can be explained by the difference in molecular tracers used and in the difference between interfer-ometric and single-dish angular resolution. CO is more abundant and widespread than SiO and has a longer gas-phase lifetime, so Maud et al. (2015) likely have greater mass sensitivity for their sample than we do for ours. However, their CO-derived outflow velocity ranges are typically narrower than those we observe with SiO by factors of $\sim 1.5$, while we have numerous small, high-velocity bullets that are more easily detected with SiO and interferometric observations. A decreased mass sensitivity but increased sensitivity to higher-velocity gas in our data would explain these comparative trends in both mass (which is velocity-independent) and momentum and energy (which are linearly- and quadratically-dependent on velocity, respectively).

We find shorter median dynamical times for our candidates ($t_{dyn} = 6,000 \pm 2,800$ years) as compared to Maud et al. (2015) ($65,000 \pm 34,000$ years). We find median, minimum, and maximum field-aggregated mass flow rates ($\dot{M}$) of $15.0^{+0.7}_{-0.7} \times 10^{-4}$ $M_\odot$ yr$^{-1}$, $2.4^{+0.1}_{-0.1} \times 10^{-4}$ $M_\odot$ yr$^{-1}$, and $130^{+10}_{-10} \times 10^{-4}$ $M_\odot$ yr$^{-1}$. Our median, minimum, and maximum field-aggregated mechanical force values ($\dot{P}$) are $3.2^{+0.02}_{-0.02} \times 10^{-2}$ $M_\odot$ km s$^{-1}$ yr$^{-1}$, $0.33^{+0.02}_{-0.02} \times 10^{-2}$ $M_\odot$ km s$^{-1}$ yr$^{-1}$, and $46^{+5}_{-5} \times 10^{-2}$ $M_\odot$ km s$^{-1}$ yr$^{-1}$. Our median, minimum, and maximum field-aggregated kinetic energy rates ($\dot{L}$) are $82^{+5}_{-5}$ $L_\odot$, $4.3^{+0.3}_{-0.3}$ $L_\odot$, and $2100^{+200}_{-200}$ $L_\odot$, respectively.

Our derived $\dot{M}$ fall within the range observed by Maud et al. ($\dot{M} = \sim 1\times10^{-5} - 1\times10^{-2}$ $M_\odot$/year 2015). Our $\dot{P}$ and $\dot{E}$ values overlap with the ranges observed by Maud et al. ($\dot{P} = \sim 7\times10^{-5} - 1\times10^{-1}$ $M_\odot$ km s$^{-1}$ year, $\dot{E} = \sim 1\times10^{-2} - 1\times10^{2}$ $L_\odot$ 2015), but our maximum values are a factor of $\sim 4$ and a factor of $\sim 20$ higher than their observed $\dot{P}$ and $\dot{E}$, respectively.

These trends are likely attributable to our higher angular resolution and our use of SiO rather than CO (both of which allow us to detect emission from smaller regions with higher velocities, i.e. smaller $t_{dyn}$), and our use of only one $t_{dyn}$ for each outflow rather than a unique $t_{dyn,i}$ for each channel.

Overall, we find that the physical properties we derive for our outflow candidates generally fall within the same ranges as those derived for similar high-mass samples at both protostellar (2,000 au) and clump ($\geq 0.1$ pc) scales. The deviations we note between our results and those in the literature are likely attributable to differences in angular resolution, molecular tracers used (CO versus SiO), and different methods of deriving dynamical times and the values that depend on them ($\dot{M}$, $\dot{P}$, $\dot{E}$).

### 4.2. Correlations Between Field-Aggregated Outflow Properties and Clump Properties

**Table 6.** Field-Aggregated Outflow Properties

| Field | $N^a_{SiO,median}$ ($10^{13}$ cm$^{-2}$) | $M^b_{out}$ (M$_\odot$) | $P_{out}$ (M$_\odot$ km s$^{-1}$) | $E_{out}$ ($10^{46}$ erg) | $\dot{M}_{out}$ ($10^{-4}$ M$_\odot$ yr$^{-1}$) | $\dot{P}_{out}$ ($10^{-2}$ M$_\odot$ km s$^{-1}$ yr$^{-1}$) | $\dot{E}_{out}$ (L$_\odot$) | $M^c_{cores,iso}$ (M$_\odot$) |
|---|---|---|---|---|---|---|---|---|
| G008.67 | 9.5 | $2.52^{+0.03}_{-0.04}$ | $52.4^{+0.6}_{-0.6}$ | $1.62^{+0.02}_{-0.02}$ | $2.4^{+0.1}_{-0.1}$ | $0.5^{+0.03}_{-0.03}$ | $14^{+1}_{-1}$ | 127 (4) |
| G010.62 | 6.5 | $8.74^{+0.09}_{-0.1}$ | $141^{+1}_{-1}$ | $4.26^{+0.05}_{-0.06}$ | $7.0^{+0.3}_{-0.3}$ | $1.42^{+0.08}_{-0.08}$ | $45^{+4}_{-4}$ | 202 (5) |
| G012.80 | 7.0 | $9.56^{+0.1}_{-0.08}$ | $184^{+2}_{-1}$ | $5.81^{+0.05}_{-0.06}$ | $16.2^{+0.8}_{-0.8}$ | $4.0^{+0.2}_{-0.2}$ | $128^{+8}_{-8}$ | 278 (5) |
| G327.29 | 10.5 | $7.54^{+0.07}_{-0.09}$ | $125^{+1}_{-1}$ | $2.79^{+0.03}_{-0.03}$ | $19.6^{+0.7}_{-0.8}$ | $3.4^{+0.1}_{-0.2}$ | $67^{+4}_{-4}$ | 1525 (5) |
| G328.25 | 20.0 | $1.40^{+0.04}_{-0.04}$ | $31.8^{+0.8}_{-0.9}$ | $1.03^{+0.03}_{-0.03}$ | $3.4^{+0.4}_{-0.4}$ | $0.8^{+0.1}_{-0.1}$ | $22^{+3}_{-3}$ | 130 (1) |
| G333.60 | 5.0 | $7.5^{+0.1}_{-0.1}$ | $128^{+2}_{-2}$ | $3.43^{+0.05}_{-0.04}$ | $15.0^{+0.7}_{-0.7}$ | $3.2^{+0.2}_{-0.2}$ | $85^{+6}_{-6}$ | 448 (5) |
| G337.92 | 6.0 | $8.7^{+0.2}_{-0.2}$ | $117^{+2}_{-2}$ | $2.30^{+0.04}_{-0.04}$ | $9.4^{+0.5}_{-0.5}$ | $1.68^{+0.08}_{-0.08}$ | $37^{+2}_{-2}$ | 507 (9) |
| G338.93 | 10.0 | $11.5^{+0.2}_{-0.5}$ | $215^{+5}_{-5}$ | $5.7^{+0.1}_{-0.1}$ | $10.5^{+0.5}_{-0.5}$ | $2.1^{+0.1}_{-0.1}$ | $51^{+3}_{-3}$ | 512 (3) |
| G351.77 | 25.0 | $11.3^{+0.6}_{-0.5}$ | $239^{+10}_{-8}$ | $8.3^{+0.4}_{-0.3}$ | $31^{+3}_{-3}$ | $6.4^{+0.5}_{-0.5}$ | $170^{+10}_{-10}$ | 515 (6) |
| G353.41 | 7.0 | $4.2^{+0.1}_{-0.1}$ | $104^{+3}_{-3}$ | $3.5^{+0.1}_{-0.1}$ | $10.3^{+0.5}_{-0.5}$ | $2.7^{+0.2}_{-0.2}$ | $82^{+5}_{-5}$ | 172 (3) |
| W43–MM1 | 10.5 | $44.6^{+0.3}_{-0.2}$ | $1039^{+9}_{-5}$ | $36.0^{+0.3}_{-0.2}$ | $89^{+3}_{-3}$ | $22.4^{+0.9}_{-0.9}$ | $680^{+30}_{-30}$ | 1683 (12) |
| W43–MM2 | 10.5 | $11.52^{+0.08}_{-0.06}$ | $238^{+1}_{-2}$ | $7.83^{+0.06}_{-0.05}$ | $15.3^{+0.5}_{-0.6}$ | $3.5^{+0.1}_{-0.1}$ | $100^{+6}_{-6}$ | 582 (4) |
| W43–MM3 | 5.5 | $4.04^{+0.05}_{-0.06}$ | $51.6^{+0.6}_{-0.6}$ | $0.9^{+0.01}_{-0.01}$ | $2.6^{+0.1}_{-0.1}$ | $0.33^{+0.02}_{-0.02}$ | $4.3^{+0.3}_{-0.3}$ | 170 (3) |
| W51–E | 35.0 | $56.6^{+0.9}_{-0.4}$ | $1550^{+20}_{-10}$ | $73.3^{+1.0}_{-1.1}$ | $130^{+10}_{-10}$ | $46^{+5}_{-5}$ | $2100^{+200}_{-200}$ | 2883 (35) |
| W51–IRS2 | 7.5 | $46.3^{+0.6}_{-0.4}$ | $990^{+10}_{-10}$ | $33.8^{+0.3}_{-0.4}$ | $79^{+5}_{-5}$ | $17^{+1}_{-1}$ | $500^{+40}_{-40}$ | 2473 (20) |

[a] The median SiO column density across all outflow candidates in the given field. Medians are calculated only from pixels meeting the significance threshold within each aperture and channel.

[b] The field-total $M_{out}$ (and $P_{out}$, $E_{out}$, etc) is the sum of the derived mass of each individual outflow candidate in the given field. Upper (lower) uncertainties are the square root of the quadrature sum of upper (lower) uncertainties for each individual candidate.

[c] The total mass in cores in each field, derived using the flux-density values for each core listed in Appendix D of Louvet et al. (2023, submitted) and assuming $T = 15$ K, $\tau << 1$, and $h\nu << kT$. Uncertainties are the square root of the quadrature sum of the uncertainties of the individual cores.

We further explore our data at the protocluster level by testing for correlations between our field-aggregated outflow properties and clump-scale properties. In particular, we explore the relationship between total outflow mass, momentum, energy, mass rate, mechanical force, and mechanical luminosity in a given protocluster and clump mass ($M_{clump}$), clump bolometric luminosity ($L_{bol}$), clump luminosity-to-mass ratio ($L_{bol}/M_{clump}$), and total mass in cores ($M_{cores,Louvet}$, $M_{cores,isotherm}$).

The clump masses are derived using the Point Process Mapping tool (PPMAP, Marsh et al. 2015) to fit a modified blackbody function to far-infrared and millimeter data for each field. We use the ALMA-IMF 1.3 mm continuum mosaics (Motte et al. 2022; Ginsburg et al. 2022), Apex Telescope Large Area Survey of the GALaxy 870 $\mu$m images (ATLASGAL, Schuller et al. 2009), and 70 $\mu$m - 500 $\mu$m Photodetector Array Camera and Spectrometer (PACS, Poglitsch et al. 2010) and Spectral and Photometric Imaging REceiver (SPIRE, Griffin et al. 2010) data from the *Herschel* telescope (Pilbratt et al. 2010). For the three fields in W43, we specifically use data from the Herschel imaging survey of OB Young Stellar Objects (HOBYS, Motte et al. 2010; Nguyen-Lương et al. 2013); for all other fields, we use data from the Herschel infrared Galactic Plane Survey (Hi-GAL, Molinari et al. 2010, 2016). For W51-IRS2 only, we replace the 250 $\mu$m Hi-GAL SPIRE data (which are saturated) with 214 $\mu$m data from the Stratospheric Observatory for Infrared Astronomy (SOFIA, Temi et al. 2014, 2018) High-resolution Airborne Wideband Camera Plus instrument (HAWC+, Harper et al. 2018). These data were obtained from SOFIA project 05_0038 (Vaillancourt 2016). We calculate the bolometric luminosities by combining the PPMAP-derived modified blackbody functions with *Spitzer* images (Werner et al. 2004): 3.6, 4.5, 5.8, and 8.0 $\mu$m data from the Infrared Array Camera (IRAC, Fazio et al. 2004) and 24 $\mu$m data from the Multiband Imaging Photometer for Spitzer (MIPS, Rieke et al. 2004). For all fields, we assume $\kappa_{300\mu m} = 0.1$ cm$^2$ g$^{-1}$ and $\beta = 1.8$, and derive background-subtracted $M_{clump}$ and $L_{clump}$ for the full 1.3 mm field of view for each field. Further details of the PPMAP procedure can be found in Dell'Ova et al. (2023, in prep).

Most fields contain only one prominent dust clump, as seen in the ATLASGAL 870 $\mu$m data. For the three fields that contain two clumps (G008.67, W43-MM3, W51-IRS2), value of $M_{clump}$ we use in this analysis is the sum of the two clumps (i.e. the mass within the entire field of view), not the value of one or the other of the ATLASGAL-detected clumps. Readers are advised that this vocabulary differs from some of the other publications in the ALMA-IMF series.

We use two separate values for total mass in cores in order to account for potential biases in the core mass derivation methods. First, we use the core masses derived in Louvet et al. (2023, submitted), who use unique temperature values for each core and take free-free emission and optical depth into account ($M_{cores,Louvet}$). This method avoids some of the uncertainties associated with assuming a uniform dust temperature and optical depth for all cores. Second, we derive our own total mass in cores using the flux-density values listed in Appendix D of Louvet et al. (2023, submitted) and a uniform temperature of 15 K assuming the optically-thin Rayleigh-Jeans approximation ($M_{cores,isotherm}$). This method gives us a more direct comparison between our results and literature samples, as this is the more common method in the literature. By using both methods, we are able to test whether our results are a result of or are robust against methodological differences. The field-aggregated outflow properties are listed in Table 6. All clump properties except $M_{cores,isotherm}$ can be found in Table 1; $M_{cores,isotherm}$ for each field are listed in Table 6.

For each pair of properties, we calculate both the Kendall $\tau$ and Spearman $\rho$ correlation coefficients and their associated $p$-values. These coefficients and $p$-values are presented as heatmaps in Figure 5. We find that most relationships are positive, but only a few correlations are significant at the $>3\sigma$ level (excluding autocorrelations and correlations between dependent properties, e.g. M and P=Mv). The aggregated outflow properties are most strongly correlated with total mass in cores ($3-5\sigma$; $\tau$ and $\rho$ values $0.6-0.9$), followed by clump mass and bolometric luminosity ($1.5-2.5\sigma$, $\tau$ and $\rho$ values $0.3 - 0.6$), and finally clump L/M ($<1\sigma$, $\tau$ and $\rho$ values $-0.01 - 0.2$).

Using the Spearman $\rho$ test, all outflow properties have $>3\sigma$ correlation with $M_{cores,isotherm}$ and all but $M_{out}$ and $E_{out}$ have $>3\sigma$ correlation with $M_{cores,Louvet}$. Using the Kendall $\tau$ test, $\dot{M}_{out}$ and $P_{out}$ have $>3\sigma$ correlation with both variations of $M_{cores}$, and $\dot{P}_{out}$ and $M_{out}$ are correlated at $>3\sigma$ with $M_{cores,isotherm}$ only. The highest correlation coefficients are between $M_{out}$ and $M_{cores,isotherm}$ in the Kendall $\tau$ test and $\dot{M}_{out}$ and

$M_{cores,isotherm}$ in the Spearman $\rho$ test ($>4.5\sigma$ for both). We discuss selected correlations in greater detail in the following subsections.

Our poor correlation with clump bolometric luminosity can be explained if both a) the field-aggregated outflow properties are the simple sum of individual outflow+protostar pairings, even though we do not associate any candidates with specific driving sources in this work, and b) the protostellar population in these fields is not purely coeval. Bontemps et al. (1996) find that outflow mechanical force ($\dot{P}$) decreases with protostellar age for their sample of low-luminosity protostars, with Class I sources following a well-defined linear relationship in log-log space and Class 0 sources lying $\sim$1 order of magnitude above this line. They suggest that the Class 0 sources follow an evolutionary track down to the best-fit line as they age into Class I sources. Bontemps et al. (1996) also find that the relationship between outflow mass rate ($\dot{M}$) and envelope mass for individual protostars ($M_{env}$, or $M_{core}$ in our data) does not appear to change with time. They suggest that individual protostars will occupy different regions of the $\dot{M} - M_{env}$ plot as they evolve (and thus both their envelope mass and accretion rates decrease), but that the relationship between the two values seems to remain approximately $\log(\dot{P}) =$ -(4.15$\pm$0.1) + (1.1$\pm$0.15)$\times$log($M_{env}$) for both Class 0 and Class I low-mass sources.

In a population of protostars which is not purely coeval, some protostars will be in the Class 0 stage, some in Class I, and some perhaps in Class II. In the Bontemps et al. (1996) interpretation, the mix of protostellar stages will not have a strong impact on correlations with envelope (core) mass for field-aggregated properties, but it will result in increased scatter in correlations with bolometric luminosity. This is consistent with what we see for our data, and therefore a plausible explanation for this difference in correlation strength; our clump sample size (15) is likely small enough to mask a correlation with luminosity (if present) due to small-number statistics. An alternate possibility is that we have additional sources of luminosity (e.g. external irradiation) which are contributing to clump luminosity but are not associated with outflows. Additional analysis in which outflow candidates are associated with specific driving sources, and bolometric luminosities for each driving source, are derived, will be needed to answer this question definitively. We therefore do not attempt to fit a relationship between our field-aggregated outflow properties and clump $L_{bol}$ or clump $L/M$ in this work. Instead, we place these results in context with the literature in § 4.2.3 and 4.2.4.

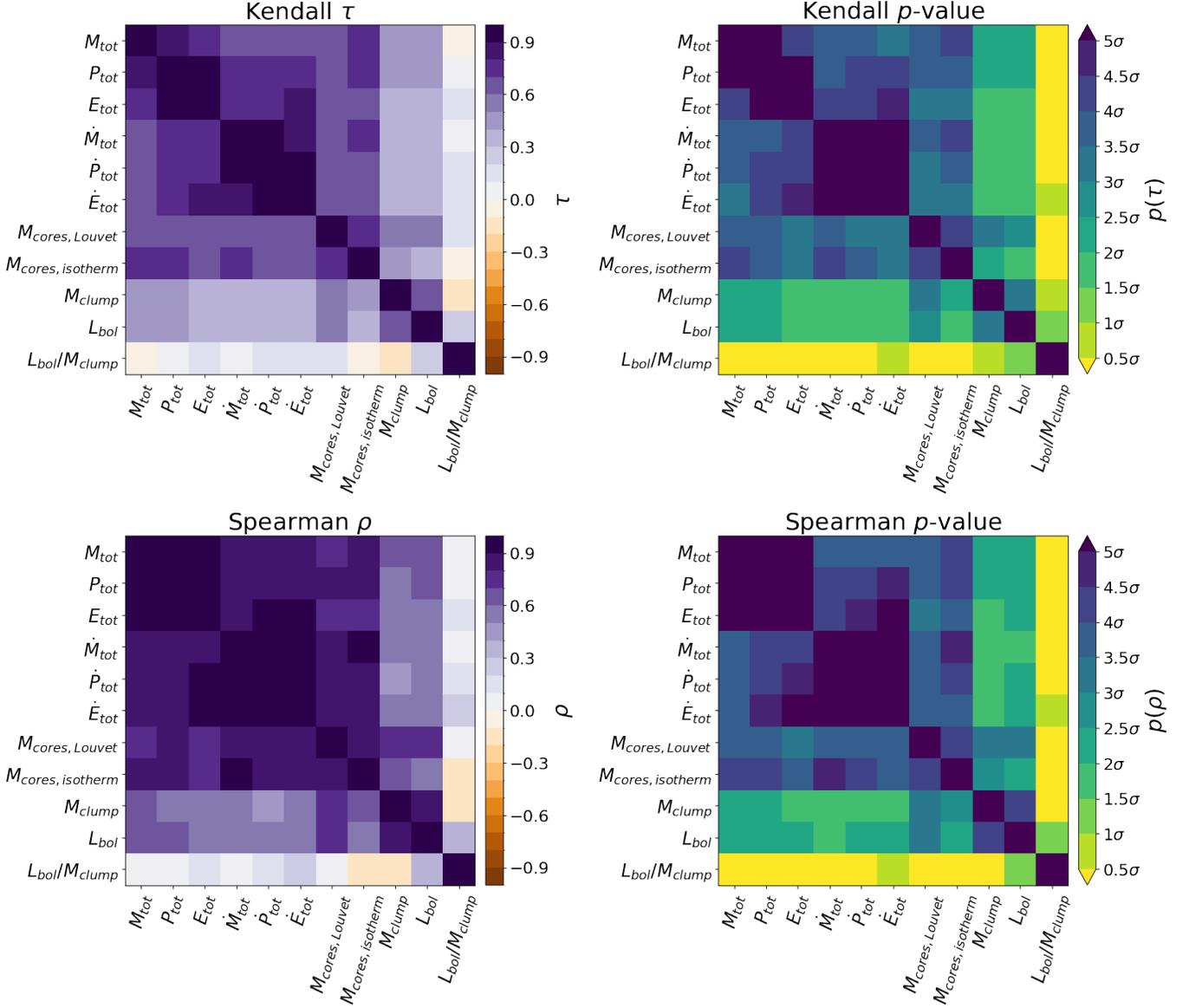

**Figure 5.** Kendall's $\tau$ (*top*) and Spearman $\rho$ (*bottom*) correlation coefficients and associated $\sigma$-values for the field-aggregated outflow properties and clump properties. Higher correlation coefficients indicate stronger positive correlations; correlation coefficients near zero indicate weak or no correlation. The $\sigma$ values show the significance of each correlation coefficient, and are derived from the $p$-values returned by the `scipy.stats packages` assuming a normal distribution. The conversions from $p$-values to $\sigma$-values are: $1\sigma = 0.32$, $2\sigma = 0.045$, $3\sigma = 0.0027$, $4\sigma = 6.3 \times 10^{-5}$, and $5\sigma = 5.7 \times 10^{-7}$. Excluding autocorrelations and correlations between coupled properties, our strongest correlations are between field-aggregated outflow properties and total mass in cores. We find no correlations above $3\sigma$ between field-aggregated outflow properties and $M_{clump}$, clump $L_{bol}$, or $M_{clump}/L_{bol}$.

### 4.2.1. *Relationship Between $M_{out}$, $M_{cores}$, and $M_{clump}$*

In this section, we consider the extent to which outflow mass is driven by the total mass in cores rather than the clump mass. Since most of the material observed in outflows is assumed to be entrained (Bally 2016), both possibilities are worth considering. We find that total outflow mass is correlated with both clump mass and total core mass in a region. We compare these correla-tions and determine that the mass in cores has a stronger effect.

We use python's `scipy.optimize.curve_fit` function to determine best-fit linear relations between $M_{out}$ and $M_{clump}$, $M_{out}$ and $M_{cores}$, and $M_{cores}$ and $M_{clump}$ in log space. We also calculate $\rho$ and $\tau$ correlation coefficients and their $p$-values for all three cases. The parameters for all best-fit lines, and $\rho$ and $\tau$ correlation coefficients and $p$-values, are shown in Table 7. We discuss only the Louvet et al. (2023, submitted) values of $M_{cores}$ in this section, but the results for the isothermally-derived $M_{cores}$ data are similar and are listed in Table 7.

Our best-fit line for $\log(M_{out})$ versus $\log(M_{clump})$ shows a positive relationship, with a slope of 0.76 and intercept of $-2$. This fit is shown in the left-hand panel of Figure 6. The $p$-values for both $\rho$ and $\tau$ correspond to a significance of $\sim 2.5\sigma$. Our fitted slope and intercept are largely consistent with previous literature values: both Li et al. (2018) and Beuther et al. (2002b) find slopes of 0.8-0.9 and an intercept of $-1$ for their fits to $\log(M_{out})$ versus $\log(M_{clump})$ for their samples of massive outflows.

Our best-fit line for $\log(M_{out})$ versus $\log(M_{cores})$ is steeper, with a slope of 1.06 and intercept of $-1.5$ (Fig. 6, middle panel, and Table 7). The correlation coefficients for $\log(M_{out})$ versus $\log(M_{cores})$ are both larger and more statistically significant than those we derive for $\log(M_{clump})$, with $\rho$ and $\tau$ $p$-values corresponding to a significance of $\sim 3.5\sigma$.

Our best-fit line to $\log(M_{cores})$ versus $\log(M_{clump})$ has a similar slope as that of $\log(M_{out})$ versus $\log(M_{clump})$, but the $\rho$ and $\tau$ $p$-values correspond to a higher degree of significance: $\sim 3.25\sigma$ for $\log(M_{cores})$ versus $\log(M_{clump})$ compared to $2.5\sigma$ for $\log(M_{out})$ versus $\log(M_{clump})$. The parameters of our best-fit line also do not agree within errors with our fit to $\log(M_{out})$ versus $\log(M_{cores})$. This suggests that the relationship between total mass in cores and clump mass is separate from the relationship between outflow mass and total mass in cores. These data and our best-fit line are shown in the right-hand panel of Figure 6.

In order to test whether and how these correlations depend on each other, we subtract the best-fit line to $\log(M_{out})$ versus $\log(M_{cores})$ from the $\log(M_{out})$ data and fit the residuals against $\log(M_{clump})$. This sequence of simple linear regression and subtraction allows us to determine how much of the correlation between $M_{out}$ and $M_{clump}$ can be explained by the relationship between $M_{out}$ and $M_{cores}$ − if the residuals still have a noticeable trend and strong $\rho$ and $\tau$ coefficients, then $M_{cores}$ cannot fully explain the correlation between $M_{out}$ and $M_{clump}$.

For each data point, we calculate $\log(M_{out,residual})$ = $\log(M_{out})$ - $f(M_{cores})$, where $f(M_{cores})$ is the y-value predicted by the best-fit line to $\log(M_{out})$ versus $\log(M_{cores})$ as shown in the middle panel of Figure 6. Then, we plot each $\log(M_{out,residual})$ value against its corresponding clump mass, and fit a line with `scipy.optimize.curve_fit`. Our results are shown in Figure 7.

Controlling for $\log(M_{out})$ versus $\log(M_{cores})$ significantly decreases the correlation with clump mass. The best-fit line is effectively flat, with a slope and intercept of zero within uncertainties. The $\tau$ and $\rho$ correlation coefficients also fall to near zero, and their $p$-values correspond to $<0.5\sigma$ significance.

We cannot invert this test − subtracting the best-fit line to $\log(M_{out})$ versus $\log(M_{clump})$ and fitting the residuals against $\log(M_{cores})$ − because $M_{clump}$ by definition includes $M_{cores}$. However, we can safely say that our current results are *not* consistent with a scenario in which clump mass directly determines total mass in outflows, with no alteration by core mass. The dominant correlation in our analysis is between outflow mass and total mass in cores, not either of the relations involving clump mass. We suggest that the total mass in cores is at least mediating the total mass in outflows to a physically significant degree, and may in fact be dominating it.

### 4.2.2. *Dominance of the Most Massive Outflow in Each Field*

The typically large distances to massive star-forming regions mean that the spatial resolution of most outflow surveys in such regions is $\geq 0.1$ pc (Beuther et al. 2002b; Maud et al. 2015; Liu et al. 2022). In most cases, this is too large to resolve individual outflows and identify their associated driving sources. Many authors therefore assume, as a first-order approximation, that the total outflow mass is dominated by the most massive individual outflow in the field, presumed to be generated by the most massive protostellar core. For example, Maud et al. (2015) use a simulated, coeval Salpeter population of protostars to test the contribution of massive protostars to the total mechanical force ($\dot{P}$) of the protocluster. They conclude that $\dot{P}$ can be entirely explained by outflows from low- and intermediate-mass protostars up to $L = 6400 L_\odot$, and dominated entirely by the massive protostars above this limit.

In order to test this assumption for our own data, we compare the mass of the most massive outflow ($M_{out,maximum}$) with total outflow mass ($M_{out,total}$) for each field. These data are plotted in Figure 8. The left-hand panel shows the most massive outflow versus total outflow mass, and the right-hand panel shows the *percentage* of total mass the most massive outflow accounts for, compared to total outflow mass in the field. There is a correlation between $M_{out,maximum}$ and $M_{out,total}$, as expected.

We find that, for our SiO-detected outflows, the most massive outflow in each field is typically responsible for only 12-30% of the measured total outflow mass, regardless of how much material is contained in outflows over-all. The most massive outflow is responsible for the majority of outflow mass ($>50\%$) in only one out of our fifteen fields. We also find no significant trend in outflow maximum-mass percentage with total outflow mass. We conclude that the total spatially-integrated mass is not dominated by the most massive outflows in our sample.

**Table 7.** Linear Regression, Kendall's $\tau$, and Spearman $\rho$ Results for log(M$_{clump}$), log(M$_{cores}$), and log(M$_{out}$)

| Independent Variable | Dependent Variable | Slope[a] | Intercept[a] | Spearman $\rho$ $\rho$, $p(\rho)$ | Kendall's Tau $\tau$, $p(\tau)$ |
|---|---|---|---|---|---|
| log(M$_{clump}$) | log(M$_{out}$) | $0.76 \pm 0.20$ | $-2.08 \pm 0.81$ | 0.62, 0.013 | 0.47, 0.016 |
| log(M$_{cores}$) (L)[c] | log(M$_{out}$) | $1.06 \pm 0.15$ | $-1.53 \pm 0.36$ | 0.80, 3.5×10$^{-4}$ | 0.67, 5.2×10$^{-4}$ |
| log(M$_{cores}$) (I)[c] | log(M$_{out}$) | $0.88 \pm 0.13$ | $-1.38 \pm 0.34$ | 0.88, 1.36×10$^{-5}$ | 0.77, 1.01×10$^{-5}$ |
| log(M$_{clump}$) | log(M$_{cores}$) (L) | $0.73 \pm 0.14$ | $-0.53 \pm 0.56$ | 0.79, 4.6×10$^{-4}$ | 0.59, 0.002 |
| log(M$_{clump}$) | log(M$_{cores}$) (I) | $0.71 \pm 0.22$ | $-0.14 \pm 0.88$ | 0.63, 0.012 | 0.47, 0.016 |
| log(M$_{clump}$) | log(M$_{out}$) - f(M$_{cores}$) (L) | $-0.01 \pm 0.13$ | $-0.05 \pm 0.53$ | 0.04, 0.89 | 0.03, 0.92 |
| log(M$_{clump}$) | log(M$_{out}$) - f(M$_{cores}$) (I) | $0.13 \pm 0.13$ | $-0.53 \pm 0.52$ | 0.31, 0.26 | 0.22, 0.28 |

[a] The slopes, intercepts, and uncertainties for each best-fit line are determined by the `scipy.optimize.curve_fit` ordinary least-squares (OLS) fitting package.

[b] The $\rho$ and $\tau$ correlation coefficients and their $p$-values are calculated with `scipy.stats.spearmanr` and `scipy.stats.kendalltau`, respectively. We convert the $p$-values to $\sigma$-values assuming a normal distribution. The conversions are: $3\sigma = 2.7\times10^{-3}$, $3.5\sigma = 4.7\times10^{-4}$, $4\sigma = 6.3\times10^{-5}$, and $4.5\sigma = 6.8\times10^{-6}$.

[c] As in Figure 5, we use two values for M$_{cores}$: those whose masses were determined with unique temperatures for each core in Louvet et al. (2023, submitted), and those assuming a temperature of 15 K for all cores (see § 4.1). The M$_{cores}$ values taken from Louvet et al. (2023, submitted) are noted with a "(L)," and those calculated assuming T = 15 k are noted with a "(I)." We discuss only the results derived from M$_{cores}$ (L) in the text, but present the M$_{cores}$ (I) results here and in Figures 6 and 7 for completeness.

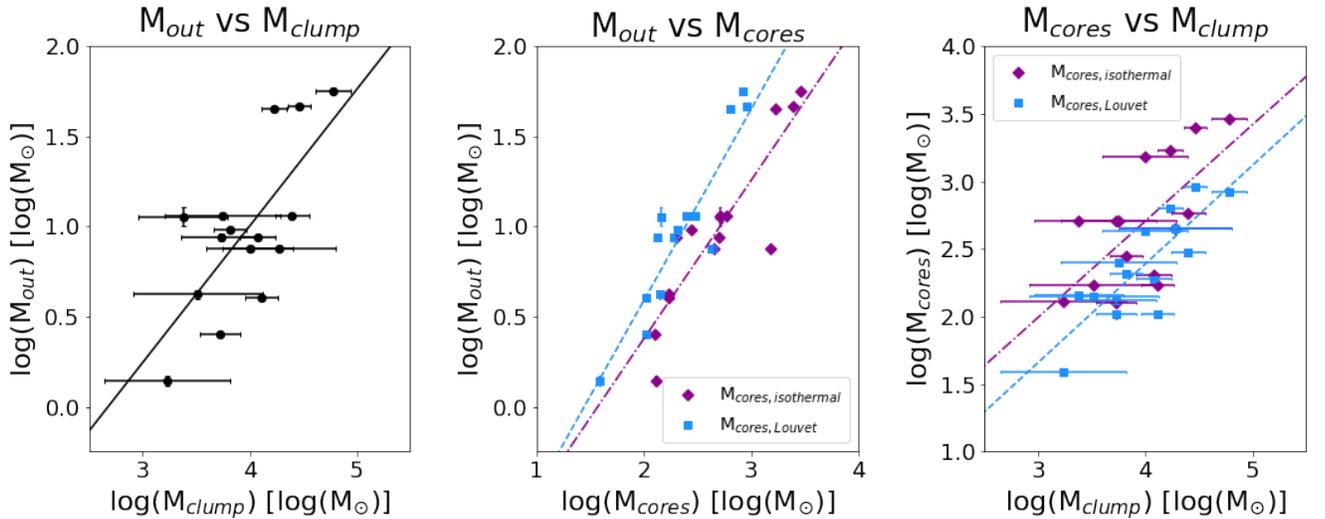

**Figure 6.** Ordinary least-squares best-fit lines to M$_{clump}$, M$_{cores}$, and M$_{out}$. In all panels, errors on both the x- and y-axes are plotted, but are too small to see in most cases except the M$_{clump}$ data. Best-fit slopes and intercepts, as well as $\rho$ and $\tau$ correlation coefficients and associated $p$-values, are listed in Table 7. *Left:* log(M$_{out}$) versus log(M$_{clump}$), with the least-squares fit shown as a solid black line. *Center:* log(M$_{out}$) versus log(M$_{cores}$) for both values of log(M$_{cores}$). The Louvet et al. data and best-fit line are shown in blue (squares and dashed line), and the isothermal data and best-fit line are shown in magenta (diamonds and dot-dashed line). We discuss only fits to the Louvet et al. M$_{cores}$ data in-text. *Right:* log(M$_{cores}$) versus log(M$_{clump}$) for both values of M$_{cores}$. Colors, symbols, and linestyles are the same as in the previous panel.

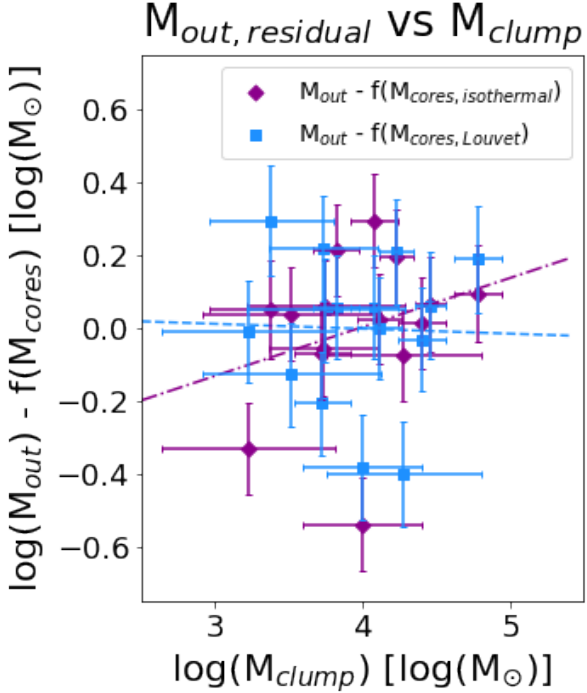

**Figure 7.** Residual outflow mass versus clump mass, after the relationship between outflow mass and total mass in cores has been subtracted from $\log(M_{out}$. Best-fit slopes and intercepts, as well as $\rho$ and $\tau$ correlation coefficients and associated $p$-values, are listed in Table 7. Residuals and best-fit lines calculated using the Louvet et al. values of $M_{cores}$ are shown in blue (squares and dashed line) and those calculated using the isothermal core masses are shown in magenta (diamonds and dot-dashed line).

We also explore this trend in outflow momentum and energy, and mass, momentum, and energy rates. The maximum values for an individual outflow in each field are shown in Table 8, along with their percentage contribution to the field total and the ID number of the candidate responsible. We find similar trends as for outflow mass - there are correlations between $P_{out,maximum}$ and $P_{out,total}$, $E_{out,maximum}$ and $E_{out,total}$, etc but no indication that the strongest outflow is responsible for $>50\%$ of each derived property. We further note that the same outflow is not always responsible for the greatest share of every property. In some fields, the same outflow does dominate all six of $M_{out}$, $P_{out}$, $E_{out}$, $\dot{M}_{out}$, $\dot{P}_{out}$, and $\dot{E}_{out}$, but in other fields, different outflows will be responsible for $M_{out,max}$ and $P_{out\,max}$, $P_{out\,max}$ and $\dot{E}_{out\,max}$, etc. Modifying our test to include the two most massive outflows brings the median $M_{out,maximum\,2}/M_{out,total}$ ratio up to 44%.

In § 3.3.2, we note the possibility of "broken" outflows, i.e. instances in which what we identify as 2 individual candidates are instead both part of the same larger outflow. This phenomenon is more common with fainter outflow candidates, not the most massive ones, but we nonetheless consider what effect this phenomenon would have on our analysis in this section. If our most massive identified outflow were instead one component of a single, larger outflow encompassing additional SiO emission, this would increase the mass of what we identify as the most massive outflow in each region by up to a factor of 2 (assuming the second component is nearly identical in mass, the largest possible case). If this were the case for every field, this would raise our typical $M_{out,maximum}/M_{out,total}$ ratios to 24-60%. Alternately, this analysis treats the red and blue lobes of bipolar outflows separately; combining the masses of each would have the same effect of at most doubling the contribution. Regardless, the typical maximum outflow contribution of $12-30\%$ (upper limit 65%) is not trivial, but is not large enough for observers to safely neglect contributions from lower-mass outflows.

### 4.2.3. *Outflow Mechanical Force versus Clump Bolometric Luminosity*

Bontemps et al. (1996) find that the relationship between outflow mechanical force ($\dot{P}$) and source bolometric luminosity for low-mass protostars evolves with time, with Class I protostars falling along a linear correlation in log-log space ($\log(\dot{P}) = -(5.6\pm0.1) + (0.9\pm0.15)\times\log(L_{bol})$) and Class 0 protostars following an evolutionary track a factor of $\sim$10 above this line. Duarte-Cabral et al. (2013) find similar results for their sample of 9 Class 0 high-mass protostars in Cygnus X, as do van der Marel et al. (2013) for their sample of 16 low-luminosity Class I sources in Ophiuchus. Maud et al. (2015) compare their sample of high-mass protoclusters to the individual protostellar samples of these previous papers, and find that their sample is reasonably well fit by the relationship derived by Bontemps et al. (1996) as well. Maud et al. (2015) additionally derive a $\dot{P}$-$L_{bol}$ relationship using only their RMS-selected data, and find a slightly shallower best-fit line of $\log(\dot{P}$ = -4.8 + 0.61$\times\log(L_{bol})$).

In Figure 9, we plot outflow mechanical force against clump bolometric luminosity for our sample. The left-hand panel shows our field-aggregated outflow mechanical force for each field as dark blue circles, and the right-hand panel shows the mechanical force of only the most powerful outflow in each field as light blue circles. In order to maintain consistency with previous literature samples, we use our inclination-corrected $\dot{P}$ values for this comparison. Correction factors for $\dot{P}$ can be found in the lower section of Table 5, and we assume a uniform inclination angle of $57.°3$ for all candidates (see Bontemps et al. 1996; Maud et al. 2015).

**Table 8.** Strongest Outflows: Absolute and Percentage Contributions, and Candidate ID Numbers[a]

| Field | $M_{max}$ | | | $P_{max}$ | | | $E_{max}$ | | |
|---|---|---|---|---|---|---|---|---|---|
| | ($M_\odot$) | (%) | (ID) | ($M_\odot$ km s$^{-1}$) | (%) | (ID) | ($10^{46}$ erg) | (%) | (ID) |
| G008.67 | 1.11 | 44 | 1 | 23.4 | 45 | 1 | 0.79 | 49 | 1 |
| G010.62 | 2.11 | 24 | 17 | 63 | 45 | 17 | 2.80 | 66 | 17 |
| G012.80 | 1.18 | 12 | 35 | 45.6 | 25 | 35 | 2.29 | 39 | 35 |
| G327.29 | 0.91 | 12 | 18 | 20.4 | 16 | 18 | 0.57 | 20 | 18 |
| G328.25 | 0.91 | 65 | 2 | 25.4 | 80 | 2 | 0.93 | 90 | 2 |
| G333.60 | 1.02 | 14 | 15 | 18.2 | 14 | 5 | 0.86 | 25 | 5 |
| G337.92 | 3.7 | 43 | 13 | 44 | 38 | 13 | 0.68 | 30 | 13 |
| G338.93 | 5.7 | 50 | 9 | 116 | 54 | 9 | 3.10 | 54 | 9 |
| G351.77 | 5.3 | 47 | 3[b] | 135 | 56 | 3[b] | 5.90 | 71 | 3[b] |
| G353.41 | 0.92 | 22 | 11 | 33 | 32 | 11 | 1.42 | 41 | 11 |
| W43−MM1 | 5.0 | 11 | 27[b] | 141 | 14 | 24 | 6.60 | 18 | 24 |
| W43−MM2 | 2.24 | 19 | 17 | 71 | 30 | 17 | 2.96 | 39 | 17 |
| W43−MM3 | 1.13 | 28 | 6 | 20 | 39 | 6 | 0.48 | 53 | 6 |
| W51−E | 10.9 | 19 | 20[b] | 538 | 35 | 20[b] | 34.4 | 47 | 20[b] |
| W51−IRS2 | 14.1 | 30 | 38[b] | 382 | 39 | 38[b] | 14.4 | 43 | 38[b] |

| Field | $\dot{M}_{max}$ | | | $\dot{P}_{max}$ | | | $\dot{E}_{max}$ | | |
|---|---|---|---|---|---|---|---|---|---|
| | ($10^{-4}$ $M_\odot$ yr$^{-1}$) | (%) | (ID) | ($M_\odot$ km s$^{-1}$ yr$^{-1}$) | (%) | (ID) | ($L_\odot$) | (%) | (ID) |
| G008.67 | 0.81 | 34 | 1 | 0.0017 | 34 | 1, 3 | 5.4 | 39 | 1 |
| G010.62 | 1.8 | 26 | 17 | 0.0053 | 37 | 17 | 20 | 44 | 12 |
| G012.80 | 3.7 | 23 | 35 | 0.014 | 35 | 35 | 60 | 47 | 35 |
| G327.29 | 2.6 | 13 | 18 | 0.0058 | 17 | 18 | 14 | 21 | 18 |
| G328.25 | 2.4 | 71 | 2 | 0.007 | 88 | 2 | 20 | 91 | 2 |
| G333.60 | 2.7 | 18 | 5 | 0.011 | 34 | 5 | 41 | 48 | 5 |
| G337.92 | 2.1 | 22 | 12 | 0.003 | 18 | 5 | 11 | 30 | 5 |
| G338.93 | 2.7 | 26 | 9 | 0.0058 | 28 | 9 | 16 | 31 | 13 |
| G351.77 | 10 | 32 | 3[b] | 0.026 | 41 | 3[b] | 100 | 59 | 3[b] |
| G353.41 | 1.9 | 18 | 11 | 0.007 | 26 | 11 | 25 | 30 | 11 |
| W43-MM1 | 9 | 10 | 18 | 0.03 | 13 | 18 | 110 | 16 | 18 |
| W43-MM2 | 3.4 | 22 | 5 | 0.011 | 31 | 5 | 38 | 38 | 5 |
| W43-MM3 | 0.7 | 27 | 10 | 0.0009 | 27 | 10 | 1.4 | 33 | 6 |
| W51-E | 56 | 43 | 19[b] | 0.22 | 48 | 19[b] | 1100 | 52 | 19[b] |
| W51-IRS2 | 24 | 30 | 38[b] | 0.06 | 35 | 38[b] | 190 | 38 | 38[b] |

[a] For each derived physical property ($M_{max}$, $P_{max}$, $E_{max}$, $\dot{M}_{max}$, $\dot{P}_{max}$, $\dot{E}_{max}$), we show the absolute value of the strongest outflow in each field (in $M_\odot$, $M_\odot$ km s$^{-1}$, etc), the fractional contribution that that outflow makes to the field-aggregated total (in %, rounded to the nearest integer), and the identification number of that candidate (ID). The same outflow is not always the dominant contributor of each property. Likewise, in one case (G008.67 $\dot{P}_{max}$) two candidates are equally strong; we list both outflow ID numbers in the relevant ID column.

[b] These candidates have known contamination from hot-core line emission within the velocity range of the SiO emission. Therefore, the physical properties reported for these candidates (especially $E$ and $\dot{E}$) should be treated as upper limits.

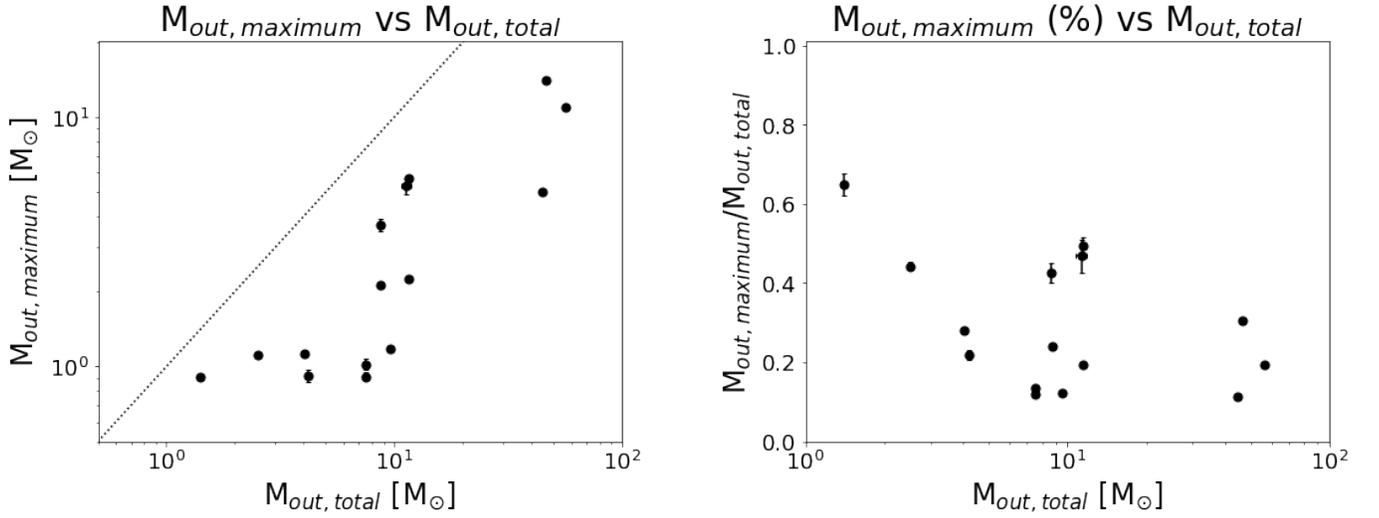

**Figure 8.** *Left:* Mass of the most massive outflow candidate in each field versus the total mass in outflows in that field. We find a positive correlation between the two properties, as expected since $M_{out,\,total}$ by definition includes $M_{out,\,maximum}$. The dotted line is the 1:1 line. *Right:* The percentage of total outflow mass that the most massive outflow is responsible for in each field ($M_{out,\,max}/M_{out,\,tot}$). We find that the most massive outflow is responsible for >50% of the total outflow mass in only one field; typically, the most massive outflow is responsible for only 15-30% of the total outflow mass. There is no correlation between $M_{out,\,max}/M_{out,\,tot}$ and $M_{out,\,tot}$.

We find that our field-aggregated mechanical force values agree well with the Bontemps et al. (1996) best-fit line, and our data extend this line up to $L > 10^6$ $L_\odot$ and $\dot{P} \sim 1.0$ $M_\odot$ km s$^{-1}$ yr$^{-1}$ for protocluster-aggregated emission. The agreement between our data and prior literature results is consistent with our earlier interpretation of these field-aggregated values being the sum of individual, well-behaved protostellar outflows with little contamination from ambient emission.

The percentage of total $\dot{P}$ that the most powerful outflow is responsible for ranges from 13% to 88%, with a median of 34%. The most powerful outflow is responsible for >50% of the total mechanical force in only one case. In other words, the most powerful outflow in a field is typically responsible for a nontrivial portion of mechanical force but not a majority, consistent with our results in § 4.2.2 for outflow mass.

We find that considering only the most powerful outflow in each field ($\dot{P}_{out,max}$) causes our data to deviate from the best-fit line of Bontemps et al. (1996), and from the larger observational trend established in the literature (Figure 9, right-hand panel). We take this as further consistency with our results in § 4.2.2 and with the broader literature. This supports the picture of the total mechanical outflow feedback in massive star-forming regions being the sum of multiple individual outflows, and which is poorly described by assuming the aggregate outflow properties are reflective of only the most massive or powerful outflow in the protocluster.

### 4.2.4. *Protocluster Outflow Properties and Clump Evolutionary State*

We find no significant trends between any outflow properties and protocluster evolutionary state, as measured by the clump luminosity-to-mass ratio ($L/M$; see Figure 5). This lack of correlation is inconsistent with models of protocluster formation in which all protostars start forming at the same time; if that were the case for our protostellar populations, we should expect outflow accretion rate and force ($\dot{M}$, $\dot{P}$) to decrease as source luminosity increases, while total clump mass remains relatively steady. Instead, we see no strong anticorrelation (or correlation of any type) between outflow properties and clump $L/M$. This suggests that the quantifiable outflow feedback in our sample is not strongly dependent on clump evolutionary state within the range of evolutionary states probed by our sample (7 $L_\odot$/$M_\odot$ $\leq$ $L/M \leq$ 79 $L_\odot$/$M_\odot$).

The lack of correlation between outflow properties and protocluster $L/M$ is consistent with the results of Liu et al. (2021) for their sample of 32 massive clumps in Infrared Dark Clouds (IRDCs), and with Liu et al. (2022) for their sample of 171 clumps in the ALMA Three-millimeter Observations of Massive Star-forming regions (ATOMS) survey. Both teams find no correlation between SiO line luminosity and clump $L/M$ for their samples, and Liu et al. (2022) interpret this as implying that SiO line luminosity and clump evolutionary state are not related.

## 5. SUMMARY & CONCLUSIONS

We have presented our first, full catalog of protostellar outflow candidates detected in SiO J=5−4 in the ALMA-IMF Large Program. In total, we detect 315 candidates across all 15 fields, with $\geq$ 3 outflow candidates in each field. We classify each outflow according to its color (red, blue, or red+blue) and likelihood (possible, likely, complex or cluster), and report approximate center positions, total velocity range, peak velocity and peak flux density of the aperture-integrated spectrum, and aperture- and velocity-integrated flux densities for each candidate. Our full catalog is presented in a machine-readable format in the online Journal and in ESCV format on Zenodo: doi:10.5281/zenodo.8350595. A representative example of the catalog is shown in Table 3.

We derive outflow column density assuming optically thin emission and an excitation temperature of $T_{ex} = 50^{+30}_{-20}$ K. To derive outflow mass, we adopt a fractional SiO abundance of $10^{-8.5}$. We derive outflow mass, momentum, and energy in each channel separately, which avoids the overestimation of outflow momentum and energy that can result from multiplying total outflow mass by the highest outflow velocity only. We then derive outflow lifetimes from the position-velocity path length for each candidate, excluding those outflow candidates classified as "complex or cluster" or which remain unresolved along their longest axis. Histograms for all of these properties are shown in Figures 3 and 4. We do not correct for (assumed) inclination angle in our derivation of outflow physical properties, but sample-wide statistics both with and without an assumed inclination angle are shown in Tables 4 and 5. We find no significant difference in typical outflow properties for red versus blue outflow candidates. A machine-readable table containing derived physical properties for all outflow candidates is available in the online Journal; a representative example is shown in Table C1.

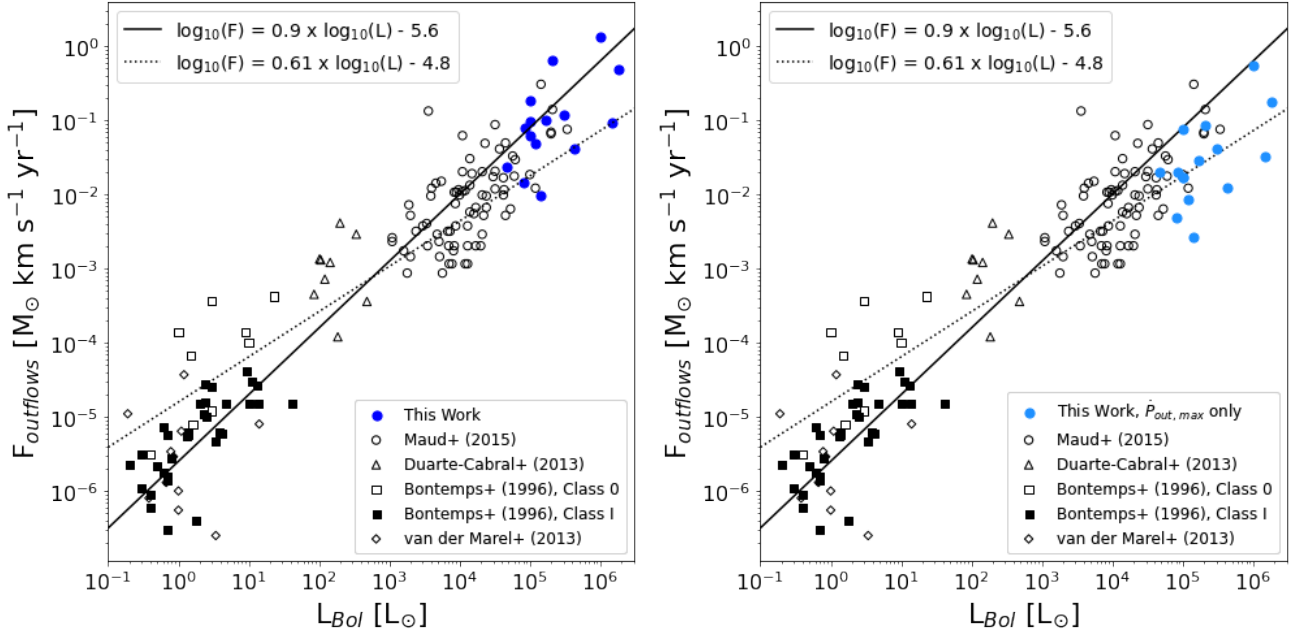

**Figure 9.** *Both Panels:* Outflow mechanical force ($\dot{P}_{out}$) versus clump bolometric luminosity (L$_{Bol}$). Open circles: massive outflows of Maud et al. (2015). Open triangles: individual massive protostars of Duarte-Cabral et al. (2013). Squares: low-mass Class 0 (open) and Class I (filled) sources of Bontemps et al. (1996). Open diamonds: low-mass sources of van der Marel et al. (2013). We overplot the best-fit lines to outflow mechanical force versus bolometric luminosity from both Maud et al. (2015) (derived using their RMS sources only; dotted line) and Bontemps et al. (1996) (derived using their low-mass sources only; solid line). *Left Panel:* This panel shows our total field-aggregated outflow mechanical force versus clump bolometric luminosity as dark blue filled circles. There is good agreement with the best-fit line of Bontemps et al. (1996) extended to higher masses, and with the overall trend of the combined literature datasets. *Right Panel:* Same as the left-hand panel, except that we show the mechanical force of only the most powerful outflow in each field as light blue filled circles.

We compare our sample to similar samples in the literature and find that our outflow properties are broadly similar. Our median SiO column density is $1 \times 10^{14}$ cm$^{-2}$, consistent with Csengeri et al. (2016) for their ATLASGAL-selected sample. Our outflow masses (range: $0.005 - 14.1$ M$_\odot$, median: $0.3 \pm 0.2$ M$_\odot$) are consistent with Lu et al. (2021) for their CMZ sample. We calculate "field-aggregated" outflow properties for each field, which are the sum of the mass, momentum, etc of the individual outflows in each field. We compare these field-aggregated values (Table 6) to Maud et al. (2015), and find that our results are broadly similar to their RMS-selected sample.

We compare our field-aggregated outflow properties to clump properties for each of our 15 fields, and test for correlations using both Kendall's $\tau$ and Spearman $\rho$ correlation tests. We find no correlations above $3\sigma$ between total outflow $M$, $P$, $E$, $\dot{M}$, $\dot{P}$, or $\dot{E}$ in a given field and clump bolometric luminosity, total clump mass, or clump $L/M$ ratio.

The lack of correlation with $L/M$ is consistent with previous literature findings for similar samples (e.g. Liu et al. 2022), which has previously been interpreted as implying overall SiO outflow properties are poorly or not at all dependent on protocluster evolutionary state. The lack of correlation with clump $L/M$ is inconsistent with models of protocluster formation in which all protostars start forming at the same time; if this were the case, we should expect to see outflow mechanical force ($\dot{P}$) decrease with clump evolutionary state, as mechanical force is known to decrease with time for individual protostars (e.g. Bontemps et al. 1996; Duarte-Cabral et al. 2013). Our best-fit line between $\log(M_{clump})$ and $\log(M_{out})$ agrees within errors with the literature values of Beuther et al. (2002b) and Li et al. (2018), even though it does not rise to $>3\sigma$ in our data.

We find that field-aggregated outflow properties are correlated at the 3-5$\sigma$ level with total mass in cores, regardless of total core-mass estimation method. We find that controlling for the relationship between total mass in cores and outflow mass strongly reduces the correlation between clump mass and outflow mass. We suggest that core mass at least mediates the total mass in outflows to a physically significant degree, and may be the primary determining factor.

Our $\log(M_{out}) - \log(M_{cores})$ correlations are intriguing because we do not associate our outflow candidates with specific driving sources. In comparing outflow mass with total mass in cores, we are comparing properties of the actively-accreting protostars only (outflow mass) with properties of the entire core population (accreting and quiescent), and still arriving at a consistent result. This consistency suggests two things. First, at the clump scale, outflows traced by SiO J=5−4 appear to be the simple sum of outflows driven by each individual protostar. Second, the tighter $\log(M_{out})−\log(M_{cores})$ correlation implies either a) a consistent fraction of protostars are accreting at any given time, or b) our 1.3 mm continuum data is more sensitive to actively-accreting protostars than to prestellar or more-evolved cores. We suggest this as a potentially fruitful avenue for future investigations.

We also examine the dominance of the most massive outflow in each field, and find that the most massive outflow is responsible for <30% of the total mass in outflows in the majority of protoclusters. Taking possible methodological bias into account, we place an upper limit on this proportion of 60%. This is not a trivial contribution, but we argue it is also not large enough for observers to safely neglect contributions from lower-mass outflows when examining field-aggregated outflow data (e.g. low spatial resolution).

Finally, we place our field-aggregated outflow mechanical force values in context with previous work by Bontemps et al. (1996), Duarte-Cabral et al. (2013), van der Marel et al. (2013), and Maud et al. (2015) examining the relationship between outflow mechanical force and source bolometric luminosity. We find that our data agree well with previous works, and extend this relationship up to $L \geq 10^6$ L$_\odot$ and $\dot{P} \geq 1.0$ M$_\odot$ km s$^{-1}$ yr$^{-1}$ using our field-aggregated data.

We thank the referee for their very constructive comments, which helped to improve this work. This paper makes use of the following ALMA data: ADS/JAO.ALMA#2017.1.01355.L, ADS/JAO.ALMA#2013.1.01365.S. ALMA is a partnership of ESO (representing its member states), NSF (USA) and NINS (Japan), together with NRC (Canada), MOST and ASIAA (Taiwan), and KASI (Republic of Korea), in cooperation with the Republic of Chile. The Joint ALMA Observatory is operated by ESO, AUI/NRAO and NAOJ. The National Radio Astronomy Observatory is a facility of the National Science Foundation operated under agreement by the Associated Universities, Inc. This project has received funding from the European Research Council (ERC) via the ERC Synergy Grant ECOGAL (grant 855130), from the French Agence Nationale de la Recherche (ANR) through the project COSMHIC (ANR-20-CE31-0009), and the French Programme National de Physique Stellaire and Physique et Chimie du Milieu Interstellaire (PNPS and PCMI) of CNRS/INSU (with INC/INP/IN2P3). The project leading to this publication has received support from ORP, that is funded by the European Union's Horizon 2020 research and innovation programme under grant agreement No 101004719 [ORP]. AG acknowledges support from the NSF under grants AST 2008101 and CAREER 2142300. AS gratefully acknowledges support by the Fondecyt Regular (project code 1220610), and ANID BASAL projects ACE210002 and FB210003. MB has received financial support from the French State in the framework of the IdEx Université de Bordeaux. PS was partially supported by a Grant-in-Aid for Scientific Research (KAKENHI Number 18H01259 and 22H01271) of the Japan Society for the Promotion of Science (JSPS). RA gratefully acknowledges support from ANID Beca Doctorado Nacional 21200897. RGM and TN acknowledge support from UNAM-PAPIIT project IN108822 and from CONACyT Ciencia de Frontera project ID 86372. TN acknowledges support from the postdoctoral fellowship program of the UNAM. YP acknowledges funding from the IDEX Université Grenoble Alpes under the Initiatives de Recherche Stratégiques (IRS) "Origine de la Masse des Étoiles dans notre Galaxie" (OMEGa). YP acknowledges funding from the European Research Council (ERC) under the European Union's Horizon 2020 research and innovation programme, for the Project "The Dawn of Organic Chemistry" (DOC), grant agreement No 741002. This research made use of NASA's Astrophysics Data System Bibliographic Services and of the SIMBAD database operated at CDS, Strasbourg, France (Wenger et al. 2000).

The following facilities are acknowledged:

*Facility:* ALMA

*Software:* s tatcont (Sánchez-Monge et al. 2018), astropy (Astropy Collaboration et al. 2022, 2018, 2013), spectral-cube (https://github.com/radio-astro-tools/spectral-cube), PV Extractor (https://github.com/radio-astro-tools/pvextractor), CASA (McMullin et al. 2007; THE CASA TEAM et al. 2022), CARTA (Comrie et al. 2021)

# APPENDIX

## A. INCIDENTAL FINDINGS IN THE DATASET

There were several incidental findings in the SiO dataset which are beyond the scope of this catalog paper. We briefly describe these in this subsection, but detailed analysis is deferred to future works.

### A.1. *Low-velocity, Narrow-line SiO Emission*

In the course of searching for protostellar outflows, we also identified a significant amount of SiO emission with no high-velocity components and no change in velocity structure with position. These regions are typically elongated in shape, similar to outflows or filaments, and their emission is within $\pm 5$ km s$^{-1}$ of the field V$_{LSR}$. Their integrated spectra are often Gaussian or sometimes triangular in shape, with linewidths $<10$ km s$^{-1}$ in all cases and $<6$ km s$^{-1}$ in most. They are found both spatially coincident with and entirely independent from high-velocity SiO emission. An isolated example of this emission is shown in Figure A1.

Similar narrow-line, low-velocity SiO emission has been previously reported in, e.g., Codella et al. (1999), Motte et al. (2007), Duarte-Cabral et al. (2014), Louvet et al. (2016), Csengeri et al. (2016), and Minh et al. (2016). The origin of low-velocity SiO emission is still a subject of some debate, but the dominant explanations at present are 1) the emission has a purely low-velocity origin, e.g., cloud-cloud collisions or slow shocks induced by gravitational collapse, or 2) the emission had a high-velocity origin initially (e.g. in protostellar outflows) and has since cooled kinematically but has not yet frozen out of the gas phase. Duarte-Cabral et al. (2014) additionally suggest that it is possible the SiO abundance was initially enhanced in a high-velocity shock but SiO is being maintained in the gas phase by low-velocity shocks alone. A detailed characterization of the low-velocity SiO emission in our sample, including tests of these possibilities, will be presented in our follow-up paper, Towner et al. 2024 (in prep).

### A.2. *Additional High-velocity Emission in G351.77*

In the field G351.77, we find additional large-scale, high-velocity emission which does not appear to trace individual protostellar outflows. This emission ranges from $-94$ to $+56$ km s$^{-1}$, and spans more than half the field of view. The morphology is bi-directional; blueshifted emission occurs predominantly north-east and north of field center and redshifted emission predominantly south-west and west, but there is little collimation in either direction. The velocity of this gas typically increases with distance from the field center, i.e. appears to exhibit Hubble flow; this trend is especially pronounced in the redshifted emission to the southwest and west. We do not include this high-velocity emission in our catalog. G351.77 is the only field with this exception. A 3-color RGB image of G351.77 is shown in Figure A2.

We suggest three primary possibilities for this emission, which will be explored in detail in a separate paper. First, this region may contain an "explosive outflow," akin to the explosive event in OMC1 (Bally et al. 2017). In OMC1, the explosion is defined by a high-velocity, roughly spherically-symmetric Hubble flow spanning $\sim 50''$ as traced by $^{12}$CO J=2-1. Here, we find lower velocities than the OMC1 explosion and less spherical symmetry, but there does still appear to be semi-coherent gas motion on the cloud scale. The second possibility is that a single massive protostar near the cloud center has recently undergone a significant episodic accretion event (e.g. Hunter et al. 2017; Caratti o Garatti et al. 2017) that ejected material at high enough velocities to produce (at least temporarily) a Hubble flow. The third possibility is that what we interpret as large-scale high-velocity emission does actually originate from individual protostellar outflows whose axes are aligned with each other. We do not favor this latter possibility at present due to the unlikelihood of both outflow axis alignment and redshift/blueshift alignment, but this scenario cannot yet be ruled out completely.

### A.3. *Bowshocks and Backsplash*

We detect arched or looping structures in the position-velocity diagrams of several of our candidates. These features are telltale signs of bowshock/backsplash in outflows colliding with the ambient medium (Bally 2016); an example from our dataset is shown in Figure A3. These features are most common in our 27 bipolar outflow candidates, but do appear in monopolar candidates as well. While we do not examine these structures in detail in this work, the dataset presented herein is one of the largest homogeneous interferometric datasets examining outflows in the literature to date. It may therefore be a useful starting point for studies of small-scale outflow physics in the future, particularly intra-outflow structure studies.

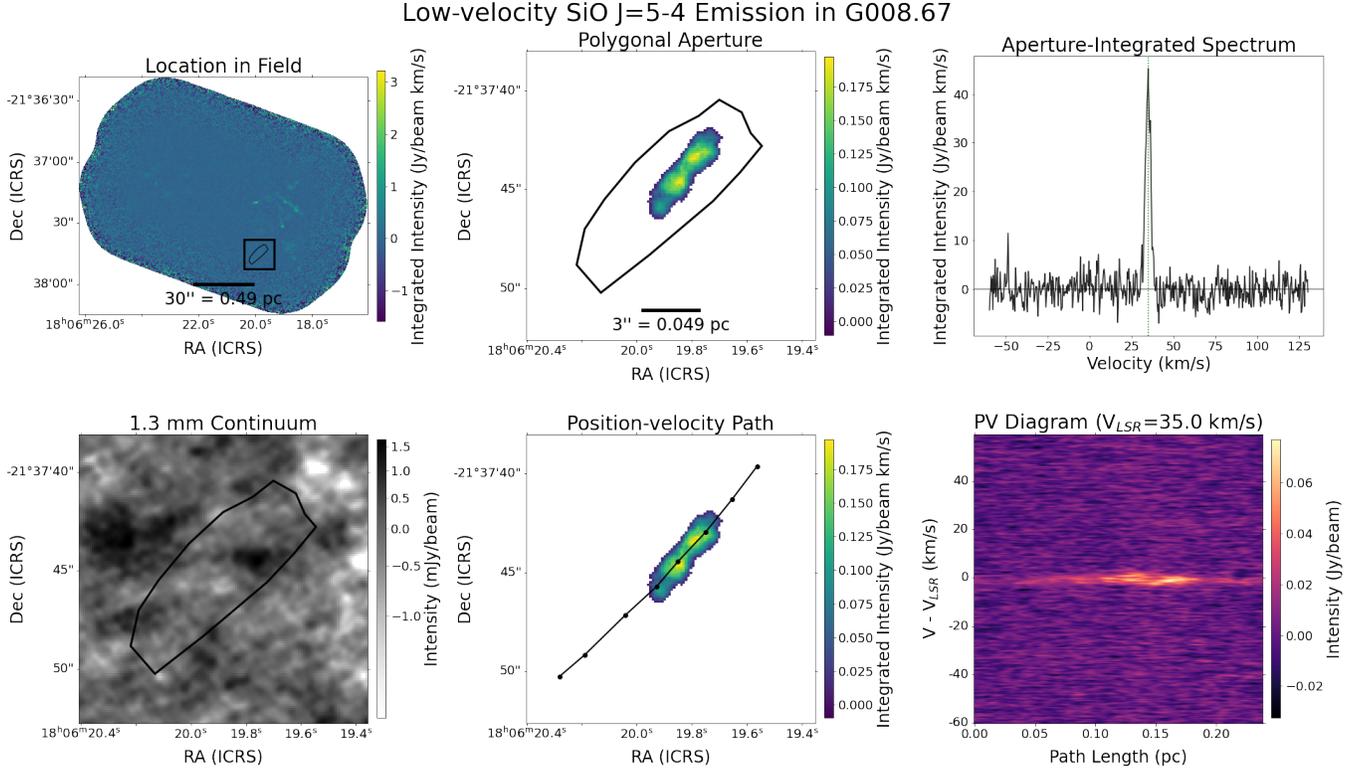

**Figure A1.** Example of isolated low-velocity emission in G008.67. This emission is not associated with an identifiable 1.3 mm continuum source, has a narrow and symmetric Gaussian line shape, and no discernible structure in its position-velocity diagram. *Top left:* Full field of view integrated-intensity (moment 0) map of G008.67 with the location of low-velocity emission highlighted in the black box. *Top center:* Zoom view of the low-velocity emission in the moment 0 map, enclosed in a polygonal aperture. *Top right:* aperture-integrated spectrum of the low-velocity emission, with field $V_{LSR}$ shown as a dotted green line. Field $V_{LSR}$ = 35.0 km s$^{-1}$. *Bottom left:* Zoom view of the 1.3 mm continuum image at the location of the low-velocity emission. *Bottom center:* Zoom view of the moment 0 map, with the position-velocity path overlaid. *Bottom right:* Position-velocity diagram for the low-velocity emission.

## B. SIO OPTICAL DEPTH, EXCITATION TEMPERATURE, AND FRACTIONAL ABUNDANCE

To derive column density, we start from the general equation for molecular column density in the optically thin approximation (see Mangum & Shirley 2015; Lu et al. 2021, equation A1):

$$N_{tot} = \frac{8\pi k_B \nu^2}{hc^3 A_{ul}} \frac{Q_{rot}}{g_J g_K g_I} \exp\left(\frac{E_u}{k_B T_{ex}}\right) \int T_B \, dv \tag{B1}$$

We then adapt this equation into a discrete form in order to calculate SiO column density in each channel individually (Eq. 1).

### B.1. *Optical Depth*

Both Eqs. B1 and 1 assume that the SiO emission is optically thin ($\tau << 1$). This is a common assumption for SiO emission (see, e.g. Lu et al. 2021), especially for higher-energy transitions such as the J=5-4 line. In some cases where direct derivations of optical depth have been done (e.g. Codella et al. 1999, using multiple $J$-transitions of SiO), SiO lines have also been directly shown to be optically thin. However, as demonstrated by the models of Gusdorf et al. (2008), SiO 5-4 may become optically thick for at least a portion of the shock lifetime in some cases (e.g. 100 yr $\lesssim t \lesssim$ 4,000 yr for initial shock parameters of 30 km s$^{-1}$ and $n_H = 10^5$ cm$^{-3}$; see Gusdorf et al. (2008) Figure 6b).

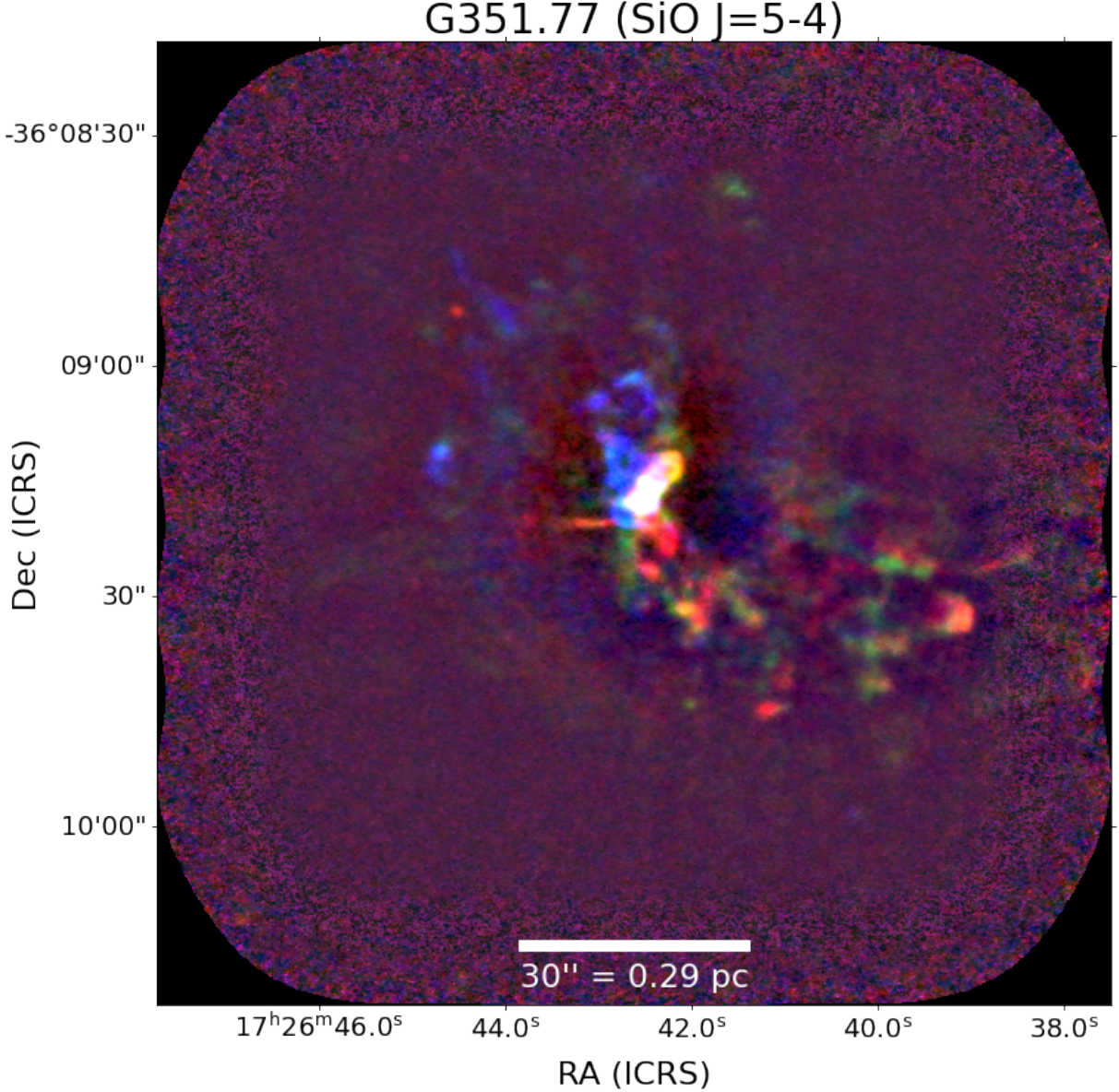

**Figure A2.** 3-color RGB figure showing the blue-shifted, ambient, and red-shifted emission in G351.77. Blue indicates SiO J=5−4 emission from −95 km s$^{-1}$ < V - V$_{LSR}$ < −5 km s$^{-1}$, green indicates SiO between −5 km s$^{-1}$ < V - V$_{LSR}$ < +5 km s$^{-1}$, and red indicates SiO between +5 km s$^{-1}$ < V - V$_{LSR}$ < +95 km s$^{-1}$.

Because we do not have multiple SiO lines available in our dataset, we cannot directly solve for $\tau_{SiO}$ for our data. We therefore assume that the lines are optically thin in all cases. This is consistent with our by-eye examination of the SiO data cubes, in which we see no clear evidence of self-absorption for any candidates.

## B.2. *Excitation Temperature*

The assumption of a single excitation temperature at all locations is unlikely to be truly physical, and so we investigate the effect of varying $T_{ex}$ on the resulting SiO column density. We find that column density is not strongly sensitive to excitation temperature in the range 30 K < $T_{ex}$ < 130 K; in this range, for a given brightness temperature, N$_{SiO}$ varies by 0.25 dex (a factor of 1.78) for our typical channel width ($\Delta v$) of 0.339 km s$^{-1}$. In Figure B1, we plot this relationship for brightness temperatures between 0.1 K and 30 K, in half-dex increments, and find the same results for all $T_B$ tested, as expected. We therefore adopt a flat excitation temperature of 50 K for all outflow candidates, with uncertainties of $-20$ K and $+30$ K. This range in temperature translates to changes of $-0.03$ dex (factor of 1.08, or 8%) and $+0.09$ dex (factor of 1.22, or 22%), respectively, in the SiO column density.

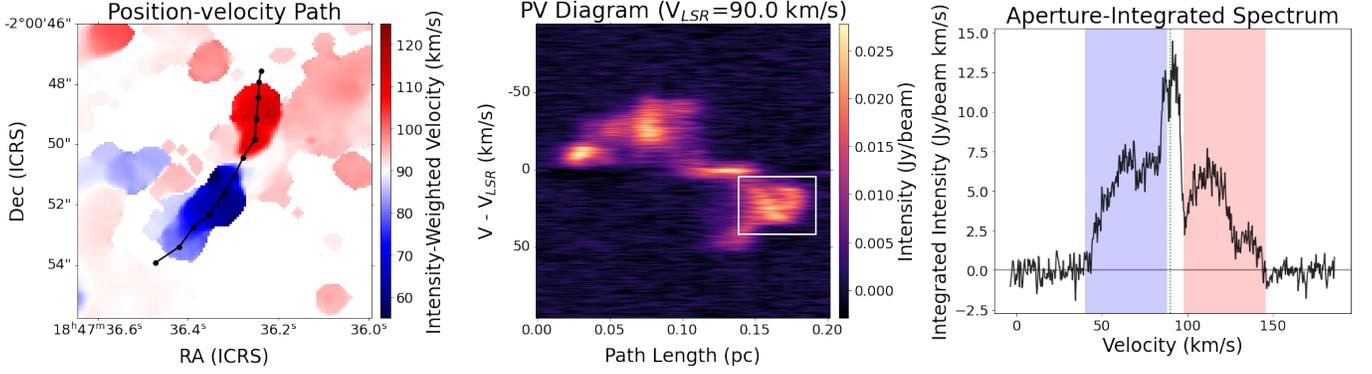

**Figure A3.** Example of structure in a position-velocity diagram indicative of bowshock/backsplash processes in W43-MM2 Candidate #8. The structure is highlighted by the white box in the middle panel. *Left:* Intensity-weighted velocity (moment 1) map of W43-MM2 Candidate #8, with the position-velocity path overlaid. Colorbar stretches from $-35$ km s$^{-1}$ $\leq$ V - V$_{LSR}$ $\leq$ $+35$ km s$^{-1}$. *Middle:* Position-velocity diagram for W43-MM2 Candidate #8. *Right:* Aperture-integrated spectrum for W43-MM2 Candidate #8, with red- and blue-shifted line wing velocity ranges highlighted in red and blue, respectively. V$_{LSR}$ is indicated by the green dotted line.

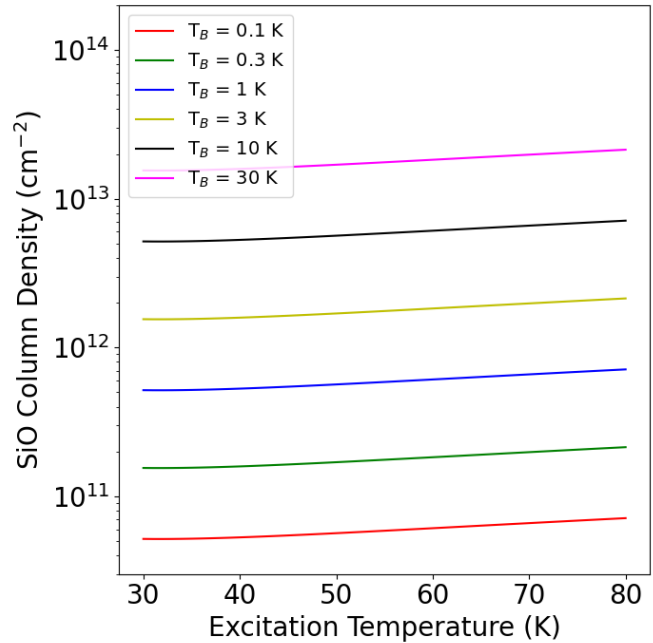

**Figure B1.** Relationship between excitation temperature ($T_{ex}$) and column density of SiO J=5–4 ($N_{SiO}$) in the optically thin approximation shown in Eq. 1 and using $\Delta$v = 0.339 km s$^{-1}$.

### B.3. *Fractional Abundance*

Eq. 3 comes from Equation A5 in Maud et al. (2015), adapted for SiO. The fractional abundance of SiO in Eq. 3 presents a particular problem, as it can vary significantly with many factors: the density of the pre-shock medium, the initial velocity of the shock, and assumptions made about the initial form of elemental Si and the subsequent astrochemical reactions which create or destroy SiO (Gusdorf et al. 2008). Indeed, Si and SiO has multiple pathways both in to and out of the gas phase in the interstellar medium, including direct release of SiO from dust-grain ice mantles or grain cores, release of elemental Si followed by a sequence of astrochemical reactions, destruction of SiO through the creation of $SiO_2$, etc. (Schilke et al. 1997) SiO fractional abundance also varies with time, depending on whether a particular parcel of gas is yet to be shocked, experiencing partial or maximum compression from the shock, or undergoing post-shock cooling (Gusdorf et al. 2008; Schilke et al. 1997). Because protostellar outflows are typically resolved phenomena, this variation with time also translates to a variation with position within an outflow (Bally 2016).

As we have no independent probe of $H_2$ column density, we cannot derive SiO fractional abundance directly. Instead, we turn to the shock-chemistry models and previous observational studies as a guide. The models of Schilke et al. (1997) suggest that SiO fractional abundance can range from a few times $10^{-11}$ to nearly $10^{-6}$, depending on post-shock time and the initial location of Si atoms (assuming a shock speed of 30 km s$^{-1}$; see their Figure 4). Likewise, the models of Gusdorf et al. (2008) find that SiO fractional abundance can range from as low as a few times $10^{-12}$ (if the initial abundance of $O_2$ ice is assumed to be negligible, and for a shock velocity of 25 km s$^{-1}$ and a preshock density of $10^5$ cm$^{-3}$) to nearly $10^{-6}$ (assuming an initial abundance of $O_2$ ice of $1.3 \times 10^{-5}$, a preshock density of $10^4$ cm$^{-3}$, and shock velocity $\geq$35 km s$^{-1}$); see their Figures 4 and C.1. In general, SiO fractional abundance in the ISM has been observed to fall range from $10^{-11}$ to $10^{-6}$, with lower values being typical in the ambient medium and values above $\sim 10^{-9}$ typical in outflows; fractional abundances at or near $10^{-6}$ have been observed toward high-velocity bullets (see Gusdorf et al. 2008; Codella et al. 1999; Schilke et al. 1997, and references therein). For their multi-species analysis of protostellar outflows in massive protoclusters in the Central Molecular Zone, Lu et al. (2021) derive fractional abundances for all species based on $H_2$ column densities, using $HC_3N$ as an anchor molecule and where the $H_2$ column densities were derived from dust continuum emission. Their derived SiO abundances range from a few times $10^{-10}$ to $10^{-8}$, with a mean value of $2.05 \times 10^{-9}$.

We therefore adopt a flat SiO-to-$H_2$ abundance ratio of $10^{-8.5}$ (or, $3.16 \times 10^{-9}$) for all candidates. This is the midpoint (in log space) of the SiO abundances in both theoretical and observational literature, and within 3% of the mean abundance (in log space) observed by Lu et al. (2021).

## C. DERIVED OUTFLOW PROPERTIES FOR EACH CANDIDATE AND FIELD

Here we present our derived properties for each individual outflow candidate, and histograms of the outflow-candidate population in each field. In Table C1, we show the first ten lines of our complete table of derived properties for each candidate. The full table can be viewed in machine-readable format in the online Journal, or in ECSV format on Zenodo: doi:10.5281/zenodo.8350595.

Our histograms are shown as a figure set in Figure C1. The example shown is for the field G008.67, and the full figure set can be viewed in the online Journal. For each histogram, the bins used are the same as in Figures 3 and 4 in order to facilitate inter-field and field-to-full-sample comparisons. For fields with $\lesssim$10 outflow candidates, the histograms are no longer smooth distributions; this is a consequence of both small-number statistics and the fact that the binning was not optimized for each field separately. Histograms are stacked, i.e., the total height of each bar represents the total number of outflow candidates (red+blue) in that bin, the red portion indicates the number of red candidates in the bin, and the blue portion indicates the number of blue candidates in the bin. Candidates classified as "complex or cluster" or which are unresolved on their longest axis (6 and 5 in total, respectively, across the full sample) are excluded from the bottom row of histograms in each figure. Readers may consult Table 3 for details as to which candidates are classified as "complex or cluster" in each individual field, and the online version of Table C1 for the list of unresolved outflow candidates.

**Table C1.** Derived Properties of Individual Outflow Candidates

| Field | ID | $N_{col, blue}$ | $N_{col, red}$ | $M_{blue}$ | $M_{red}$ | $M_{tot}$ | $P_{blue}$ | $P_{red}$ | $P_{tot}$ | $E_{blue}$ | $E_{red}$ | $E_{tot}$ |
|---|---|---|---|---|---|---|---|---|---|---|---|---|
| | | (cm$^{-2}$) | (cm$^{-2}$) | (M$_\odot$) | (M$_\odot$) | (M$_\odot$) | (km M$_\odot$ s$^{-1}$) | (km M$_\odot$ s$^{-1}$) | (km M$_\odot$ s$^{-1}$) | (erg) | (erg) | (erg) |
| G008.67 | 1 | $2\times10^{14}$ | $0.9\times10^{14}$ | 0.81 | 0.302 | 1.11 | 15.1 | 8.3 | 23.4 | $3.8\times10^{45}$ | $4.1\times10^{45}$ | $7.9\times10^{45}$ |
| G008.67 | 2 | $\cdots$ | $2.1\times10^{14}$ | $\cdots$ | 0.59 | 0.59 | $\cdots$ | 12.0 | 12.0 | $\cdots$ | $2.98\times10^{45}$ | $2.98\times10^{45}$ |
| G008.67 | 3 | $1.1\times10^{14}$ | $\cdots$ | 0.51 | $\cdots$ | 0.51 | 13.7 | $\cdots$ | 13.7 | $4.9\times10^{45}$ | $\cdots$ | $4.9\times10^{45}$ |
| G008.67 | 4 | $0.19\times10^{14}$ | $0.16\times10^{14}$ | 0.0197 | 0.008 | 0.028 | 0.22 | 0.06 | 0.28 | $2.6\times10^{43}$ | $4.7\times10^{42}$ | $3.1\times10^{43}$ |
| G008.67 | 5 | $0.6\times10^{14}$ | $\cdots$ | 0.029 | $\cdots$ | 0.029 | 0.23 | $\cdots$ | 0.23 | $1.8\times10^{43}$ | $\cdots$ | $1.8\times10^{43}$ |
| G008.67 | 6 | $\cdots$ | $1\times10^{14}$ | $\cdots$ | 0.251 | 0.251 | $\cdots$ | 2.8 | 2.8 | $\cdots$ | $3.7\times10^{44}$ | $3.7\times10^{44}$ |
| G010.62 | 1 | $1\times10^{14}$ | $\cdots$ | 0.2 | $\cdots$ | 0.2 | 1.62 | $\cdots$ | 1.62 | $1.47\times10^{44}$ | $\cdots$ | $1.47\times10^{44}$ |
| G010.62 | 2 | $1\times10^{14}$ | $\cdots$ | 0.28 | $\cdots$ | 0.28 | 1.9 | $\cdots$ | 1.9 | $1.46\times10^{44}$ | $\cdots$ | $1.46\times10^{44}$ |
| G010.62 | 3 | $0.4\times10^{14}$ | $\cdots$ | 0.141 | $\cdots$ | 0.141 | 1.57 | $\cdots$ | 1.57 | $2.11\times10^{44}$ | $\cdots$ | $2.11\times10^{44}$ |
| G010.62 | 4 | $2\times10^{14}$ | $\cdots$ | 0.24 | $\cdots$ | 0.24 | 2.1 | $\cdots$ | 2.1 | $2\times10^{44}$ | $\cdots$ | $2\times10^{44}$ |

| $t_{dyn}$ | $\dot{M}_{blue}$ | $\dot{M}_{red}$ | $\dot{M}_{tot}$ | $\dot{P}_{blue}$ | $\dot{P}_{red}$ | $\dot{P}_{tot}$ | $\dot{E}_{blue}$ | $\dot{E}_{red}$ | $\dot{E}_{tot}$ |
|---|---|---|---|---|---|---|---|---|---|
| (yr) | (M$_\odot$ yr$^{-1}$) | (M$_\odot$ yr$^{-1}$) | (M$_\odot$ yr$^{-1}$) | (km M$_\odot$ s$^{-1}$ yr$^{-1}$) | (km M$_\odot$ s$^{-1}$ yr$^{-1}$) | (km M$_\odot$ s$^{-1}$ yr$^{-1}$) | (L$_\odot$) | (L$_\odot$) | (L$_\odot$) |
| 10000 | $5.1\times10^{-5}$ | $3\times10^{-5}$ | $8.1\times10^{-5}$ | 0.0009 | 0.00083 | 0.0017 | 1.9 | 3.5 | 5.4 |
| 9000.0 | $\cdots$ | $6.6\times10^{-5}$ | $6.6\times10^{-5}$ | $\cdots$ | 0.0013 | 0.0013 | $\cdots$ | 2.8 | 2.8 |
| 8000.0 | $6.4\times10^{-5}$ | $\cdots$ | $6.4\times10^{-5}$ | 0.0017 | $\cdots$ | 0.0017 | 5.3 | $\cdots$ | 5.3 |
| 7000.0 | $4.5\times10^{-6}$ | $1.1\times10^{-6}$ | $5.6\times10^{-6}$ | $5\times10^{-5}$ | $9\times10^{-6}$ | $5.9\times10^{-5}$ | 0.049 | 0.0058 | 0.055 |
| 7000.0 | $4.1\times10^{-6}$ | $\cdots$ | $4.1\times10^{-6}$ | $3.3\times10^{-5}$ | $\cdots$ | $3.3\times10^{-5}$ | 0.021 | $\cdots$ | 0.021 |
| 14000.0 | $\cdots$ | $1.8\times10^{-5}$ | $1.8\times10^{-5}$ | $\cdots$ | 0.0002 | 0.0002 | $\cdots$ | 0.21 | 0.21 |
| 9000.0 | $2.2\times10^{-5}$ | $\cdots$ | $2.2\times10^{-5}$ | 0.00018 | $\cdots$ | 0.00018 | 0.13 | $\cdots$ | 0.13 |
| 9000.0 | $3.1\times10^{-5}$ | $\cdots$ | $3.1\times10^{-5}$ | 0.00021 | $\cdots$ | 0.00021 | 0.14 | $\cdots$ | 0.14 |
| 5700.0 | $2.5\times10^{-5}$ | $\cdots$ | $2.5\times10^{-5}$ | 0.00028 | $\cdots$ | 0.00028 | 0.31 | $\cdots$ | 0.31 |
| 13000.0 | $1.8\times10^{-5}$ | $\cdots$ | $1.8\times10^{-5}$ | 0.00016 | $\cdots$ | 0.00016 | 0.13 | $\cdots$ | 0.13 |

[a] The full table is available in the online Journal in machine-readable format, and in ECSV format on Zenodo at doi:10.5281/zenodo.8350595.
The full table includes upper- and lower-bound uncertainties for each column.

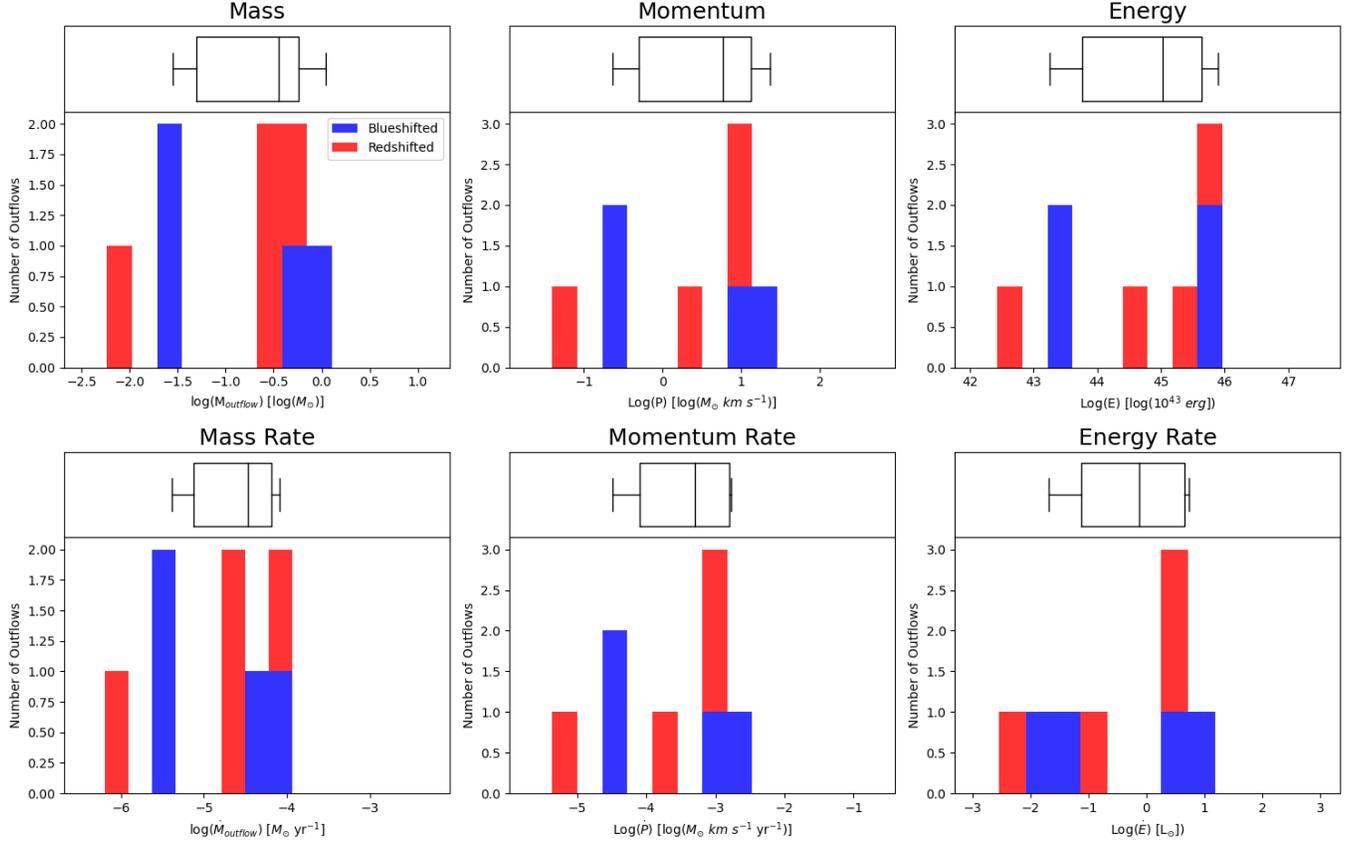

**Figure C1.** The distribution of mass, momentum, energy, mass rate, momentum rate, and energy rate for all outflow candidates in field G008.67. Candidates either classified as "complex or cluster" or unresolved along their longest axis are excluded from the rates plots (bottom row). Red bars indicate redshifted outflows, and blue bars indicate blueshifted outflows. The histogram is stacked. Histogram bins are the same as in Figures 3 and 4 for consistency of comparison between fields. Box-and-whisker plots have the same meaning as in Figures 3 and 4, but for the candidate population in G008.67 only. The complete figure set (15 images) is available in the online Journal.